\documentclass[12pt,preprint]{aastex}
\shorttitle{Runaway migration and the formation of hot Jupiters}
\shortauthors{F.\ S.\ Masset \& J.\ C.\ B.\ Papaloizou}
\begin{document}
\title{Runaway migration and the formation of hot Jupiters}
\author{F.\ S.\ Masset\altaffilmark{1}}
\affil{Service d'Astrophysique, CE-Saclay, 91191 Gif/Yvette Cedex, France}
\email{fmasset@cea.fr}
\and
\author{J.\ C.\ B.\ Papaloizou}
\affil{Astronomy Unit,  School of Mathematical Sciences,
           Queen Mary University of London, Mile End Road, London E1 4NS,
           United Kingdom}
\email{J.C.B.Papaloizou@qmul.ac.uk}
\altaffiltext{1}{Send offprint requests to F.\ S.\ Masset: fmasset@cea.fr}
\begin{abstract}
  We evaluate the coorbital corotation torque on a migrating
  protoplanet. The coorbital torque is assumed to come from orbit
  crossing fluid elements which exchange angular momentum with the
  planet when they execute a U-turn at the end of horseshoe
  streamlines. When the planet migrates inward, the fluid elements of
  the inner disk undergo one such exchange as they pass to the outer
  disk. The angular momentum they gain is removed from the planet, and
  this corresponds to a negative contribution to the corotation
  torque, which scales with the drift rate. In addition, the material
  trapped in the coorbital region drifts radially with the planet
  giving a positive contribution to the corotation torque, which also
  scales with the drift rate.  These two contributions do not cancel
  out if the coorbital region is depleted, in which case there is a
  net corotation torque which scales with the drift rate and the mass
  deficit in the coorbital region, and which has same sign as the
  drift rate.  This leads to a positive feedback on the migrating
  planet. In particular, if the coorbital mass deficit is larger than
  the planet mass, the migration rate undergoes a runaway which can
  vary the protoplanet semi-major axis by $50$~\% over a few tens of
  orbits. This can happen only if the planet mass is sufficient to
  create a dip or gap in its surrounding region, and if the
  surrounding disk mass is larger than the planet mass. This typically
  corresponds to planet masses in the sub-Saturnian to Jovian mass
  range embedded in massive protoplanetary disks. Runaway migration is
  a good candidate to account for the orbital characteristics of close
  orbiting giant planets, most of which have sub-Jovian masses.  These
  are known to cluster at short periods whereas planets of greater
  than two Jovian masses are rare at short periods indicating a
  different type of migration process operated for the two classes of
  object.  Further, we show that in the runaway regime, migration can
  be directed outwards, which makes this regime potentially rich in a
  variety of important effects in shaping a planetary system during
  the last stages of its formation.
\end{abstract}
\section{Introduction\label{sec:intro}}
The study of the tidally induced migration of protoplanets embedded in
protoplanetary disks has received renewed attention in the last few
years following the discovery of extrasolar giant planets (hereafter
EGPs). It is in particular the best candidate to explain the short
period EGPs (the so-called hot Jupiters) which are likely to have
begun to form further out in the disk and migrated radially inwards.

When the planet mass is small (i.e. when its Hill radius is much
smaller than the disk thickness), the migration rate can be evaluated
using linear analysis, and is shown to be proportional to the planet
mass, to the disk surface density, and inversely proportional to the
square of the disk aspect ratio \citep{w97}. The linear regime is
often called the type~I regime. It corresponds to a fast migration
rate, although recent estimates \citep{miyo99,tanaka,m02} show that
the linear analytical estimate assuming a flat two dimensional disk
has to be reduced by a factor $2$--$3$ or more in a more realistic
calculation which accounts for the disk vertical structure and a
possible non-saturation of the corotation torque if the disk is
viscous enough.  Migration in the type I regime is nevertheless still
too fast, in the sense that the migration timescale it yields is
shorter than the build-up timescale of a giant protocore \citep[see
e.\ g.\ ][]{p99}.  We shall not address this issue here but rather
assume that a giant planet embryo can form in the disk at a distance
$r\geq 1$~AU, and with a mass $M>M_{\rm crit}$, where $M_{\rm
  crit}\sim 15$~$M_\oplus$ is the critical mass above which rapid gas
accretion begins.

When this embryo mass is large enough, it enters another well studied
migration regime, called type~II migration \cite{w97}. In this regime,
the protoplanet has a mass sufficient to open a gap in the disk, which
is therefore split into an inner disk and an outer disk. The
protoplanet then finds itself locked into the disk viscous evolution
drifting inwards with it \citep{lp86,t98}.  As the protoplanet
undergoes type~II migration towards the central object, it may accrete
the surrounding nebula material. The accretion rate scales with the
mass accretion rate onto the central object $\dot M_p\sim
3\pi\nu\Sigma,$ where $\Sigma$ is the disk surface density.  Here one
assumes that the processes at work in the disk which contribute to the
angular momentum exchange between neighboring rings can be adequately
modeled by a phenomenological kinematic viscosity $\nu$. On the other
hand, the type~II migration timescale is of the order of the disk
viscous timescale $\tau_{\rm mig}^{\rm II}\sim\frac{r^2}{3\nu}$. The
maximum mass that a giant protoplanet can accrete on its way to the
central object should be therefore of the order of $M_p\sim \dot
M_p\tau_{\rm mig}^{\rm II}\sim\pi r^2\Sigma$, that is of the order of
the disk mass interior to the planet starting distance. Noticeably
this mass does not depend on the disk viscosity. If the planet does
not migrate all the way to the central object before the disk is
dispersed, then because more time is spent at larger radii, it is most
likely to be left with a semi-major axis larger than the typical one
for hot Jupiters \citep[$0.05$--$0.2$~AU, see][]{t02}.  This is
consistent with the observed paucity of planets with masses exceeding
two Jovian masses at short periods \citep{zm02}.  Note too that
planets undergoing type II migration should tend to have higher masses
at shorter periods.  This is contrary to the observed trend.
Furthermore, as the planet mass grows, it becomes eventually larger
than the surrounding disk mass, and its migration rate tends to drop,
as the disk cannot remove enough angular momentum from it. This has
been investigated by \citet{i99}. In this case, the amount of time
necessary to bring the planet to a close orbit can be considerably
larger than the disk viscous timescale, and can even exceed the disk
lifetime. This migration slowing can be seen in the simulations of
\citet{n00}.  It occurs {\em soon} after the planet has entered its
type II migration regime.

From the above considerations, it is questionable whether the orbital
characteristics of most close orbiters can be accounted for by type~II
migration driven by the evolution of the disk. Furthermore the vast
majority of these have sub-Jovian masses (here one excepts
{\objectname[]{Tau~Boo}} and {\objectname[]{GJ~86}}, which have large
masses and which may have had a different origin).  Depending on the
physical conditions of the protoplanetary disks in which they formed,
they may not have fulfilled the gap opening criteria \citep[][ and
references therein]{lp86}, with the consequence that they may have
been involved in a migration regime intermediate between type~I and
type~II.

This transitional regime has received little attention. \citet{w97}
and \citet{wh89} have worked out the feedback on the migration rate
due to the nebula surface density profile perturbation under the
action of the protoplanet Lindblad torques.  They introduced the
concept of an inertial limit, that is the maximum mass of a
protoplanet that can undergo steady state migration. It was suggested
that masses above the inertial limit lead to a gap opening and to
type~II migration.  In their analysis, \citet{wh89} neglected the
coorbital dynamics and the corotation torque it implies on the
migrating planet. The purpose of this work is to give an evaluation of
the corotation torque for a migrating planet, and to analyze its
consequences on migration. We define the notation in
section~\ref{sec:notation}, we present an appropriate expression for
the corotation torque for a planet held on a fixed circular orbit in
section~\ref{sec:corotfix}, we then derive the corotation torque for a
migrating planet in section~\ref{sec:derivation}, and illustrate its
properties using customized numerical simulations in
section~\ref{sec:numres}.  As we investigate the intermediate regime
between type~I and type~II migration, for which the disk response is
affected by non-linear effects, the Hill radius and the disk thickness
have comparable orders of magnitude. The regime of interest thus
involves mildly embedded protoplanets. We assume that it can be safely
studied through two dimensional flat disk calculations provided a
reasonable value is adopted for the gravitational potential smoothing
length.  In section~\ref{sec:discussion} we demonstrate the existence
of runaway migration and relate the condition for its occurrence to
the analytic discussion in section~\ref{sec:derivation}.  Finally in
section~\ref{sec:summary} we summarize our results and we discuss
their application to EGPs.

\section{Notation}
\label{sec:notation}
We adopt a cylindrical coordinate system $(r,\theta,z)$ centered on
the primary, the origin of which corotates with the planet which is
located at azimuth $\theta=0$.  We consider a thin gaseous disk with
mid-plane at $z=0$, with surface density $\Sigma(r)$, orbiting a
central point-like object of mass $M_*$. The associated Keplerian
frequency is $\Omega_K(r)=\sqrt{GM_*/r^3}$.  In the unperturbed disk
the orbital frequency is $\Omega_0(r)$, which is usually slightly
smaller than $\Omega_K(r)$ because the central acceleration is not
only compensated for by the centrifugal acceleration, but also by a
radial pressure gradient. The disk thickness is~$H(r)$ and the disk
aspect ratio is~$h(r)=H(r)/r$. In this disk we consider an embedded
planet with mass $M_p$ (we note $q=M_p/M_*\ll 1$) and semi-major axis
$a$.  Its orbital frequency is $\Omega_p=\Omega_K(a)$. We shall
restrict ourselves to the case where its eccentricity~$e$ can be
neglected. We now write the angular frequency as
$\Omega(r,\theta)=\Omega_0(r)+v/r$, and we denote by $u(r,\theta)$ the
radial velocity of the fluid element at $(r,\theta)$.

\section{Corotation torque for a planet on a fixed circular orbit}
\label{sec:corotfix}
The link between the corotation torque and the horseshoe orbit drag
has been indicated by \citet{wlpi91,wlpi92}. A torque estimate for a
planet on a fixed circular orbit embedded in a viscous disk has been
given by \citet{m01} who considered a steady flow as seen in the
planet frame. We here derive an expression for the corotation torque
in a disk with very small viscosity.  Then the specific vorticity is
conserved along a streamline \citep[ and references therein]{bk01}.

We assume a steady flow in the planet frame. As we assume that the
viscosity is low, we can write a Bernoulli invariant along a
streamline~:
\begin{equation}
\label{eq:bernou}
J=\frac{u^2+r^2(\Omega-\Omega_p)^2}{2}+\phi_{\rm eff}+\eta,
\end{equation}
where $\eta$ is the fluid specific enthalpy and $\phi_{\rm
  eff}=\phi-r^2\Omega_p^2/2$, with $\phi$ being the gravitational
potential.  The Bernoulli invariant is a useful label of the librating
streamlines in the horseshoe region. The corotation torque expression,
following \citet{wlpi91}, can be obtained by summing the contribution
to the torque due to individual fluid elements over the horseshoe
annular slab. The positive contribution coming from the outer fluid
elements caught up by the planet, when they execute a U-turn in front
of the latter, reads:
\begin{equation}
\label{eq:tqp}
  \Gamma^+=\int_{r_c}^{r_c+x_s}\Sigma(r_+)r_+(\Omega_p-\Omega)[j(r_+)-j(r_-)]dr_+,
\end{equation}
where $x_s$ is the horseshoe zone half-width, $j(r)=r^2\Omega$ is the
specific angular momentum, $r_c$ is the corotation radius, and where
we add an index $+$/ $-$ to any quantity to refer to its value on the
outer/ inner part of its horseshoe streamline.  The gradient of the
Bernoulli invariant is linked to the flow vorticity \citep[see e.\ g.\ 
][]{fr97}.  If we denote by $\omega$ the vertical component of the
flow vorticity in the inertial frame, then we have:
\begin{equation}
\label{eq:djdr}
\frac{\partial J}{\partial r}=r\omega(\Omega-\Omega_p),
\end{equation}
where we use the fact that $r\Omega^2=\partial(\phi+\eta)/\partial r$,
and where we evaluate this expression sufficiently far from the planet
so that we can assume $u=0$ and so that we can neglect the dependency
of $J$ upon $\theta$.  This allows one to transform Eq.~(\ref{eq:tqp})
as:
\begin{equation}
  \Gamma^+=-\int_{J(r_c)}^{J(r_c+x_s)}\frac{\Sigma(J)}{\omega}\gamma(J)dJ,
\end{equation}
where we note $\gamma(J)=j(r_+)-j(r_-)$ the specific angular momentum
drop of a fluid element as it switches from the outer horseshoe leg to
the inner one. If $w=\Sigma/\omega$ is the inverse of the specific
vorticity, then:
\begin{eqnarray}
\label{eq:tqp_u}
\Gamma^+&=&16|A_p|B_p^2a\int_0^{x_s}w(x)x^2dx,
\end{eqnarray}
where $A_p=(1/2)r\partial\Omega/\partial r$ and
$B_p=(1/2r)\partial(r^2\Omega)/\partial r$ are respectively the first
and second Oort's constants, evaluated at the planet orbit. Here we
assume that the planet mass is small enough that we can consider that
a fluid element on an outer horseshoe leg at $r=x+a$ is mapped on the
inner leg with radius $r=a-x$, we have developed $\gamma(x)$ to first
order in $x$ [$\gamma(x)=4B_pax$], and we have assumed $r_c=a$.  A
similar treatment for the inner horseshoe leg yields the torque
exerted on the protoplanet by the fluid elements which catch the
planet up and are promoted to higher specific angular momentum orbits:
\begin{eqnarray}
\label{eq:tqm_u}
\Gamma^-&=&16|A_p|B_p^2a\int_0^{-x_s}w(x)x^2dx.
\end{eqnarray}
It should be noted that the corotation torque, in the non-linear
regime (i.\ e. for $x_s$ finite) should include all the fluid elements
which corotate, in average, with the planet, that is to say all the
fluid elements which librate in the corotating frame. This includes
not only the horseshoe streamlines, but also the circumplanetary disk
(corresponding to the closed streamlines interior to the Roche lobe).
We find it more convenient to consider only the horseshoe drag exerted
on the system \{planet + circumplanetary disk\}, considered as a
whole, and hereafter the planet should be understood as this system.

\section{Corotation torque on a migrating planet}
\label{sec:derivation}
We shall now use Eq.~(\ref{eq:tqp_u}) and~(\ref{eq:tqm_u}) to evaluate
the torque on a migrating object. We separate the orbital timescale
$O(\Omega^{-1})$ (which is also the horseshoe U-turn timescale from
one leg to another) from the horseshoe libration timescale
$O(\Omega^{-1}a/x_s)$, which is much longer. In particular, we neglect
migration over the orbital timescale, while we consider it over the
libration timescale.  {We say that migration is slow whenever the
  libration time of an outermost fluid element close to the separatrix
  between librating and circulating streamlines is short compared to
  the migration time across the horseshoe zone half width. As the
  former quantity is $2\pi a/(|A_p|x_s)$ and the latter $x_s/|\dot
  a|$, the condition for slow migration reads:
\begin{equation}
\label{eq:slow}
|\dot a|\ll\frac{|A_p|x_s^2}{\pi a},
\end{equation}
which is also, for a Keplerian disk:
\begin{equation}
\label{eq:slow2}
|\dot a| \ll \frac 32\cdot  \frac{x_s}{\tau_{\rm orb}}\cdot \frac{x_s}{a},
\end{equation}
where $\tau_{\rm orb}=2\pi/\Omega$ is the orbital time scale.}  We
assume that Eqs.~(\ref{eq:tqp_u}) and~(\ref{eq:tqm_u}) are still
valid, provided care is taken about the evaluation of $w(x).$ In
particular, as the planet migrates, the ``impact parameter'' $x=r-a$
of a fluid element varies as its azimuth varies.  We therefore have to
consider the outer incident fluid elements at an azimuth
$\theta_R\rlap{\lower4pt\hbox{$\sim$}}\hbox{$>$}0$.  This azimuth
needs to be small enough so that no significant radial drift occurs
between the fluid element and the planet before the close encounter,
and large enough so that the close encounter has not begun yet.
Similarly, the azimuth at which the incident inner fluid elements need
to be considered has to be
$\theta_L\rlap{\lower4pt\hbox{$\sim$}}\hbox{$<$}2\pi$.  We use the
index $R$/ $L$ for the quantities relating to the close encounters
originating from $\theta_R\rlap{\lower4pt\hbox{$\sim$}}\hbox{$>$}0$ /
$\theta_L\rlap{\lower4pt\hbox{$\sim$}}\hbox{$<$}2\pi$).  The positive
part of the corotation torque is given by:
\begin{equation}
\label{eq:tqwp}
\Gamma_+=16|A_p|B_p^2a\int_0^{x_s}w_R^+(x)x^2dx,
\end{equation}
while the negative part of the corotation torque is given by:
\begin{equation}
\label{eq:tqwm}
\Gamma_-=16|A_p|B_p^2a\int_0^{-x_s}w_L^-(x)x^2dx.
\end{equation}
Note that we assume the value of $x_s$ to be the same as in the
non-migrating case. Although this needs to be reconsidered for a
significant drift rate, this is surely true as long as migration is
slow (with the meaning defined earlier in this section). The numerical
simulations presented in the next section will be used to check this
assumption.  The sum of Eqs.~(\ref{eq:tqwp}) and~(\ref{eq:tqwm}) can
be transformed so that values of $w_L(x)$ and $w_R(x)$ on the same
side of the orbit are considered. Care has to be taken about the sign
of $\dot a$ before making this transformation.  We shall assume
hereafter that $\dot a<0$. Hence:
\begin{equation}
\label{eq:tqinter}
\Gamma=16|A_p|B_p^2a\int_{-x_s}^0[w_R^-(x)-w_L^-(x)]x^2dx,
\end{equation}
where we use the fact that $w_R^+(x)=w_R^-(-x)$ for any $x$ in
$[-x_s,+x_s]$, since all of the fluid elements which execute a right
U-turn are trapped in the coorbital region (see also
Figs.~\ref{fig:appendix} and~\ref{fig:appendix2} in
appendix~\ref{apA}).

\begin{figure}
  \plotone{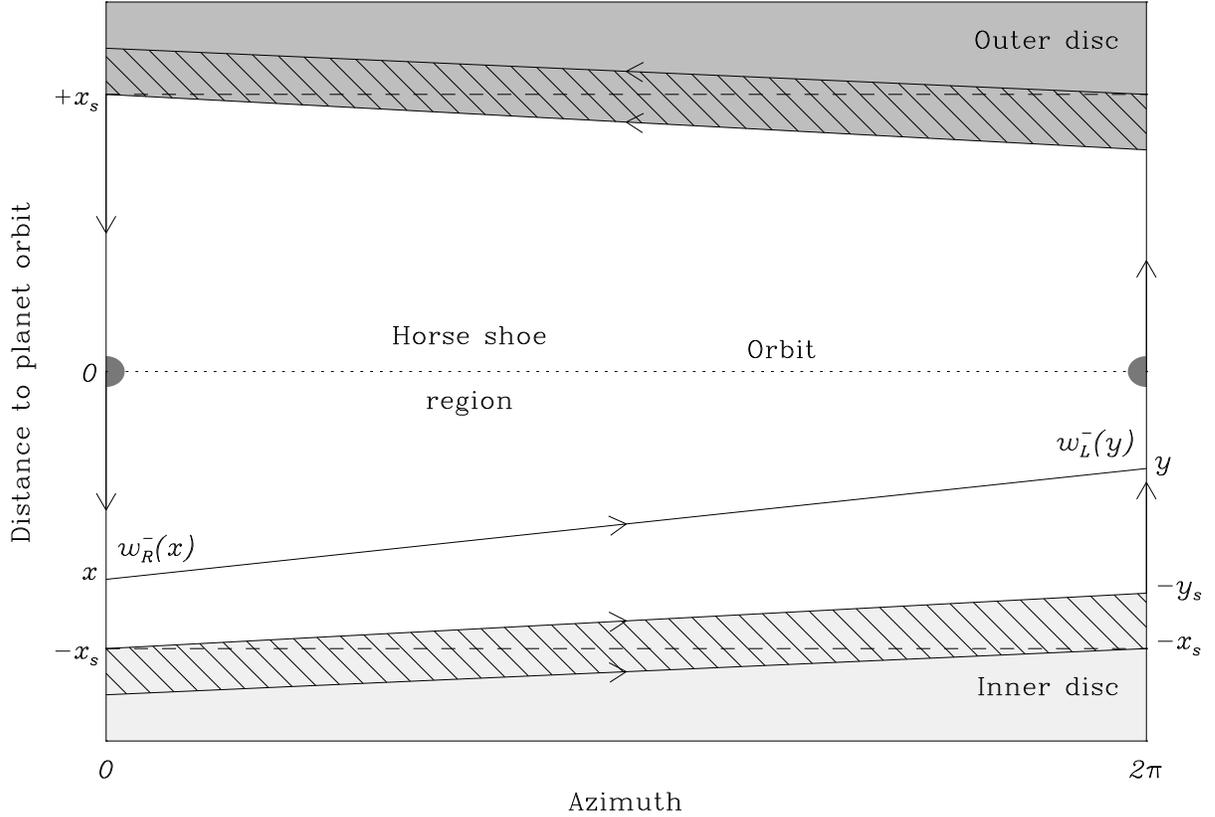}
\caption{\label{fig:interp}Sketch of a fluid element path for a situation with $\dot a<0$. The planet
  is orbiting to the right.  It is located at
  $\theta=0$~(mod.~$2\pi$).  The fluid elements paths are represented
  in a $(\theta, x=r-a)$ plane. The separatrices are represented by
  dashed lines.  The distance to the orbit of the fluid element is
  initially $x$ (just after a $R$-close encounter) and is $y$ after a
  half libration time (just before a $L$-close encounter). The bottom
  hashed zone represents the material from the inner disk that will
  have crossed the inner separatrix before the next close encounter.
  Therefore this material participates once in the corotation torque,
  and then flows out of the horseshoe region at the outer separatrix
  and circulates in the outer disk.}
\end{figure}

We shall now temporarily assume a steady migration case ($\ddot a=0$).
We call $f:x\mapsto y$ the mapping of a fluid element, between two
close encounters with the planet. This notation is illustrated in
Fig.~\ref{fig:interp}.  Consider a fluid element initially located at
the distance $x$ of the orbit just after a close encounter. Its
distance to the orbit just before the next close encounter is:
\begin{equation}
y = f(x) = x+ \frac{\pi a}{|A_p|x} \dot a,
\end{equation}
where we assume that $y\simeq x$ (i.\ e. slow migration).  Within this
approximation, the reciprocal map $f^{-1}$ reads:
\begin{equation}
x = f^{-1}(y)=y- \frac{\pi a}{|A_p|y} \dot a.
\end{equation}

Because of the conservation of specific vorticity, one can write:
$w_R^-(x) =w_L^-(y)$, for any $x$ in $[-x_s,0]$.
Eq.~(\ref{eq:tqinter}) therefore reads:
\begin{eqnarray}
\int_{-x_s}^0w_R^-(x)x^2dx&=&
\int_{-y_s}^0w_L^-(y)\left(y-\frac{\pi a\dot a}{|A_p|y}\right)^2\nonumber\\
&&\times\left(1+\frac{\pi a\dot a}{|A_p|y^2}\right)dy,
\label{eq:inter1}
\end{eqnarray}
where $y_s=f(x_s)$, and where we have taken $x=y=0$ as an upper limit
for both integrals, since the contribution of material close to the
orbit is vanishingly small for slow migration\footnote{More precisely,
  the material at low $|x|$ is not mapped onto $y$, and one should
  write $-x_T$ as an upper limit of the L.H.S. integral of
  Eq.~(\ref{eq:inter1}) instead of $0$, where $-x_T$ is defined by
  $f(-x_T)=0$. One finds $x_T^2=8\pi a\dot a/(3\Omega)$. The
  contribution of material in the $[-x_T,0]$ region is therefore $\sim
  (1/3)w_R(0)x_T^3$, and scales therefore as $\dot a^{3/2}$, whereas
  the effect of migration on the corotation torque is in $\dot a$. The
  relative contribution of this innermost material is therefore
  vanishingly small for slow migration, and it is safe to write $0$ as
  an upper limit for the L.H.S. integral.}.  Expanding the integrand
to first order in $a\dot a/(|A_p|y^2)$, one can write the corotation
torque given by Eq.~(\ref{eq:tqinter}) as:
\begin{eqnarray}
\Gamma_{\rm sm}&=16|A_p|B_p^2a&\left[\int_{-y_s}^0w_L^-(y)y^2\left(1
-\frac{\pi a\dot a}{|A_p|y^2}\right)dy\right.\nonumber\\
&&\left.-\int_{-x_s}^0w_L^-(x)x^2dx\right],\label{eq:sm1}
\end{eqnarray}
where the $\rm sm$ index stands for ``steady migration''.
Eq.~(\ref{eq:sm1}) can be transformed into:
\begin{eqnarray}
\Gamma_{\rm sm}&=16|A_p|B_p^2a&\left[-\int_{-y_s}^0w_L^-(y)
\frac{\pi a\dot a}{|A_p|}dy\right.\nonumber\\
&&-\left.\int_{-x_s}^{-y_s}w_L^-(x)x^2dx\right].
\label{eq:tqmod}
\end{eqnarray}
The first integral (over $y$, from $-y_s$ to $0$) corresponds to the
material which librates in the horseshoe region (white trapezoidal
area in Fig.~\ref{fig:interp}). The corresponding torque expression
is:
\begin{eqnarray}
\Gamma_1&=&-16\pi B_p^2a^2\dot a\int_{-y_s}^0w_L^-(y)dy\nonumber\\
&=&-2B_pa\dot a M_{\rm coorb},
\label{eq:gamma1}
\end{eqnarray}
where $M_{\rm coorb}$, the ``vorticity weighted coorbital mass'', is
defined as:
\begin{eqnarray}
M_{\rm coorb}&=&4\pi aB_p\int_{-y_s}^0\frac{\Sigma_L(y)}{B(y)}dy\nonumber\\
&=&4\pi aB_p\int_{-x_s}^0\frac{\Sigma_R(x)}{B(x)}dx.\label{eq:coorbmass}
\end{eqnarray}
This component $\Gamma_1$ of the torque arises because the librating
fluid elements migrate radially with the planet and have to lose
specific angular momentum at the same rate as this latter.  For the
case of an inward migration, the torque exerted on this region by the
planet is negative. It exerts therefore a positive torque on the
planet, and thus has a negative feed back on migration. A similar
conclusion applies for the case of an outwards migration.

The second integral in Eq.~(\ref{eq:tqmod}) can be evaluated assuming
$y_s\approx x_s$ (i.\ e. slow migration), in which case it reduces to:
\begin{eqnarray}
\Gamma_2&=&-16|A_p|^2B_p^2a(x_s-y_s)w_L(-x_s)x_s^2\nonumber\\
&=&8\pi B_p^2a^2x_s\dot a\frac{\Sigma(-x_s)}{B(-x_s)}.
\label{eq:gamma2}
\end{eqnarray}
As this torque corresponds to the integral over $-x_s<y<-y_s$, it
comes from the fluid elements of the hashed area of
Fig.~\ref{fig:interp}. These fluid elements are promoted to higher
specific angular momentum trajectories after their (unique) close
encounter with the planet. For the case of inwards migration, they
therefore contribute negatively to the corotation torque and hence
exert a positive feed back on migration.

When one adds Eqs.~(\ref{eq:gamma1}) and~(\ref{eq:gamma2}), one gets
the following torque expression:
\begin{equation}
\label{eq:tqexp_one}
\Gamma=2B_pa\,\delta m\, \dot a,
\end{equation}

where we introduce the ``vorticity weighted coorbital mass deficit''
defined as:
\begin{eqnarray}
\delta m&=&\displaystyle8\pi a\left[x_sw_R(-x_s)-\int_{-x_s}^0w_R(x)dx\right]B_p\nonumber\\
&=&\displaystyle 4\pi a\left[x_s\frac{\Sigma_R(-x_s)}{B(-x_s)}-
\int_{-x_s}^0\frac{\Sigma_R(x)}{B(x)}dx\right]B_p.
\label{eq:cmd}
\end{eqnarray}
If one neglects the radial variation of $B(x)$ across the horseshoe
region, the coorbital mass deficit appears as the mass difference
between the mass of the horseshoe region, if the material in it had
everywhere the surface density that it has at the inner separatrix,
with its actual mass. As the coorbital region is generally depleted,
this coorbital mass deficit is positive.

One can then get an estimate of the torque for a non-steady migration
using the following simple argument: if $\dot a$ varies, the mass flux
across the upstream separatrix varies, and this will have an impact on
the torque exerted on the planet when the deficit or excess of
in-flowing mass w.r.t. a steady state situation undergoes a close
encounter with the planet, i.\ e. after a characteristic time equal to
$\tau_{\rm lag}=\tau(x_s)/2$, where $\tau=(\pi a/|A_p|)|x|^{-1}$ is
half the libration time, i.\ e. the time a fluid element at a distance
$|x|$ from the orbit needs to go from azimuth $\theta_R\approx 0$ to
azimuth $\theta_L\approx2\pi$ .  Therefore $\tau_{\rm lag}$ is the
time that a fluid element opposite the planet needs to drift to attain
conjunction with it.  One can therefore write:
\begin{eqnarray}
\label{eq:tqexp}
\Gamma&=&\Gamma_{\rm sm}[\dot a(t-\tau_{\rm lag})]\nonumber\\
&=&\Gamma_{\rm sm}(\dot a)-\tau_{\rm lag}\ddot a\frac{d\Gamma_{\rm sm}}{d\dot a}\nonumber\\
&=&2B_p\,\delta m\, a\dot a-\frac{\pi a^2B_p}{|A_p|x_s}\,\delta m\,\ddot a.
\end{eqnarray}
An exact method to evaluate the $\ddot a$ term in the slow migration
limit, when one knows the specific vorticity profile across the
horseshoe region, is provided in appendix~\ref{apA}.  It can be noted
that this torque expression cancels out for a planet held on a fixed
circular orbit ($\dot a= \ddot a = 0$). This is expected as we have
used $w_R^-(-x)=w_R^+(x)$, i.\ e.  that the specific vorticity is
conserved along a streamline. Hence in the absence of migration, the
torque is saturated.  We could have used another dependency of
$w_R^-(-x)$ on $w_R^+(x)$, involving a radial gradient of specific
vorticity, which would have led to a constant term which we would
interpret as the ``static'' part of the corotation torque. However, as
our concern is to capture the planet drift effects on the torque, this
would not have brought further insight regarding the runaway process
we aim at characterizing.

We note $\Delta\Gamma_{\rm LR}$ the other torque that is applied to
the planet, which corresponds to non-librating (i.\ e., circulating)
material, which we assume to correspond to the differential Lindblad
torque. Assuming that migration occurs with a negligible eccentricity,
we have:
\begin{equation}
2B_paM_p\dot a=\Gamma+\Delta\Gamma_{\rm LR}.
\end{equation}
This takes also the following form:
\begin{equation}
\label{eq:adot1}
2B_pa(M_p-\delta m)\dot a=\Delta\Gamma_{\rm LR}-
\frac{\pi a^2\,\delta mB_p}{x_s|A_p|}\ddot a,
\end{equation}
which also takes the following form in the Keplerian case:
\begin{equation}
\label{eq:adot2}
a\frac{\Omega_p}{2}(M_p-\delta m)\dot a=\Delta\Gamma_{\rm LR}-
\frac{\pi a^2\,\delta m}{3x_s}\ddot a.
\end{equation}
Assuming that the $\dot a$ variations, if any, occur on a time
interval short enough to consider $a$ as a constant, one gets two
different behaviors from Eq.~(\ref{eq:adot1}):
\begin{enumerate}
\item $M_p>\delta m$: the coorbital mass deficit is smaller than the
  planet mass, which is the case either for sufficiently low planet
  masses or for a large planet mass, when the planet mass is
  comparable to or larger than the surrounding disk. In that case the
  coorbital mass deficit cannot become larger than the planet mass.
  Then the corresponding homogeneous equation indicates that
  disturbances to $a$ are damped on the short timescale
\begin{equation}
\label{eq:estab}
\tau_d\simeq\tau_{\rm orb}\frac{a}{x_s}\frac{\delta m}{M_p-\delta m}.
\end{equation}
One can therefore discard transient behavior, retaining only the
standard first order ODE for migration:
\begin{equation}
\label{eq:ode1}
a\frac{\Omega_p}{2}(M_p-\delta m)\dot a=\Delta\Gamma_{\rm LR},
\end{equation}
where the only difference with the usual expression is that we replace
the planet mass alone $M_p$ by the ``planet effective mass'' $m_{\rm
  eff}=M_p-\delta m$.
\item $M_p < \delta m$: the planet mass is large enough to open a
  significant dip in the disk, and this latter is substantial enough
  for the coorbital mass deficit to be larger than the planet mass. In
  that case the homogeneous ODE associated to~Eq.~(\ref{eq:adot1})
  indicates that small perturbations to $a$ are exponentially growing
  on a timescale $\tau_d$, which although depending on the exact value
  of $|M_p-\delta m|/M_p$, is of the order of a few tens of orbital
  periods (assuming, as is reasonable for a mildly embedded object in
  a typical protoplanetary disk, that $x_s\sim 0.1$). The assumptions
  that we made to derive of Eq.~(\ref{eq:tqexp}), namely migration
  slow enough that the dip profile drifts instantaneously with the
  planet, and slow enough that the horseshoe zone crossing time is
  much larger than the libration time, rapidly break down, and one can
  only say at this point that this runaway regime is extremely fast,
  and that it occurs for $\delta m>M_p$.  The actual behavior of a
  protoplanet in this regime has to be assessed through numerical
  simulations.  Since the ultimate sign of $\ddot a$ depends on the
  initial values of $a$ and $\dot a$, runaway can in principle occur
  outwards under specific initial conditions which need to be
  specified.
\end{enumerate}

\section{Numerical simulations}
\label{sec:numres}

\subsection{Code description}
We performed a series of dedicated numerical simulations to test the
runaway regime and the validity of~Eq.~(\ref{eq:ode1}). The code that
we used has already been described elsewhere \citep[see e.\ g.\ 
][]{n00}. As this code is an Eulerian grid-based code, it must fulfill
the Courant condition on the timestep to ensure numerical stability.
An improved algorithm resulting in a less demanding CFL condition,
with the average azimuthal velocity at each radius subtracted out
\citep{m00,m00b} was used in order to increase the time step and speed
up the code.  The grid corotates with the guiding center of the planet
osculating orbit. As a result, the planet motion with respect to the
grid is slow, and mainly corresponds to a radial drift.  Our rotating
frame is angularly accelerated. The corresponding acceleration, that
is $\vec r \times \dot{\Omega}_p\vec{e_z}$, is applied in much the
same way as \citet{k98} handles the Coriolis acceleration in a
rotating frame, so as to enforce angular momentum conservation.  As
these dedicated runs involve as accurate as possible a torque
evaluation, we used non-reflecting boundary conditions so as to
eliminate any reflected wave (which can bring back to the planet the
angular momentum previously removed from it) and we used an initial
profile with uniform specific vorticity (i.\ e. with $\Sigma\propto
r^{-3/2}$, which leads to a constant drift rate $\dot a$ up to the
center, if one only considers the differential Lindblad torque).  Our
mesh consists of $450$ sectors equally spaced in azimuth, divided
radially in $143$ zones, the successive radii of zone boundaries being
in geometric progression.

\subsection{Units and Setup}
\label{sec:setup}
Our unit of mass is the central object mass $M_*$, our unit of length
is the initial planet semi-major axis $a_0$ and our unit of time is
$\Omega_K(a_0)^{-1}$. In this system of units, the gravitational
constant is therefore $G=1$. The mesh outer boundary lies at $2.5a_0$
and the inner boundary at $0.4a_0$.  We use a uniform aspect ratio
disk with $h(r)\equiv 0.03$. The resolution of our mesh is barely
sufficient to accurately describe the differential Lindblad torque
acting on the planet.  On the other hand, it is enough for a proper
description of coorbital effects \citep{m02}.  Our planet mass is
$M_p=3\cdot 10^{-4}$ (which corresponds to a Saturn mass planet if the
central star has a solar mass). We used a number of disk surface
density profiles
\begin{equation}
\Sigma(r)=S_n\cdot 10^{-4}\cdot r^{-3/2}
\end{equation}
with the following values $S_0=0.5$, $S_1=1$, $S_2=1.5$, $S_3=2$,
$S_4=3$, $S_5=5$, $S_6=8$, $S_7=12$ and $S_8=20$. We used a uniform
kinematic viscosity $\nu = 10^{-5}$ throughout the disk.  The planet
was initially held on a fixed circular orbit for $477$~orbits
($t=3000$) in order to give it a sufficient time to open a dip/gap
around its orbit. This creates a depression of the specific vorticity
profile which can lead to the effect we described in section 2.  It
should be noted that the coorbital mass deficit that arises during the
first $477$~orbits just scales with the disk surface density, that is
to say: $\delta m\propto \Sigma_0$, where $\Sigma_0=\Sigma(a_0)$.
Eq.~(\ref{eq:ode1}) can therefore be written as:
\begin{equation}
\label{eq:d1}
(M_p-K\Sigma_0)\dot a=K'\Sigma_0
\end{equation}
where the constants $K$ and $K'$ both depend on $M_p$ and the disk
parameters, but this dependency does not need to be considered here as
we only vary the disk surface density in these runs.
Eq.~(\ref{eq:d1}) can be written as:
\begin{equation}
\label{eq:check}
\dot a^{-1}=A(\Sigma_0^{-1}-\Sigma_{\rm crit}^{-1})
\end{equation}
where we introduce the critical surface density $\Sigma_{\rm crit}$
for migration runaway, with $A$ being a proportionality constant which
is simply related to the differential Lindblad torque, as one can see
by letting $\Sigma_0\rightarrow 0$.  The novelty here is that the
migration rate below the runaway limit grows faster than linearly with
the disk surface density. Our simulations aim at testing this
super-linearity by checking whether the drift rate
fulfills~Eq.~(\ref{eq:check}) or not.  When it does, we determine the
critical disk surface density for runaway and determine whether we
indeed get a runaway for larger surface densities.  We also check that
the critical surface density so determined is consistent with
coorbital mass deficit estimates.

\subsection{Smoothing issues}
The protoplanet potential in the runs presented here is smoothed using
a softening parameter $\varepsilon =\eta H$, where $H$ is the disk
thickness and $\eta = 0.6$. The results turn out to be sensitive to
the value of the softening parameter because the horseshoe zone width
crucially depends on it. Lowering the softening parameter moves the
separatrices away from the orbit. As a result, the coorbital mass
deficit is increased, and the critical surface density for runaway is
reduced.  However, sensitivity to the softening parameter in a
numerical simulation does not necessarily imply that the region within
the Roche lobe matters, but rather that coorbital processes matter, as
their effectiveness depends strongly on the horseshoe zone width which
is linked to it \citep[see][ \S~5 for details]{m02}. Note that the
discussion of the softening parameter by \citet{m02} is valid only for
the case when the specific vorticity has a linear dependence on the
distance to corotation (in which case the corotation torque scales as
$x_s^4$).  Here the specific vorticity profile has a depleted, more
complex profile, and the analysis is no longer valid. We used a
softening parameter comparable to the one derived by \citet{m02},
which was found to give reasonable results for the case of mildly
embedded planets. It should be kept in mind that if a smaller
softening parameter is used, the critical surface density for runaway
that we shall discuss later would be even lower than what we found. In
that sense the extent of the runaway regime that we shall delineate
later in this work can be considered as a conservative estimate.

\subsection{Results}
\label{sec:results_run_1}
\begin{figure}
  \plotone{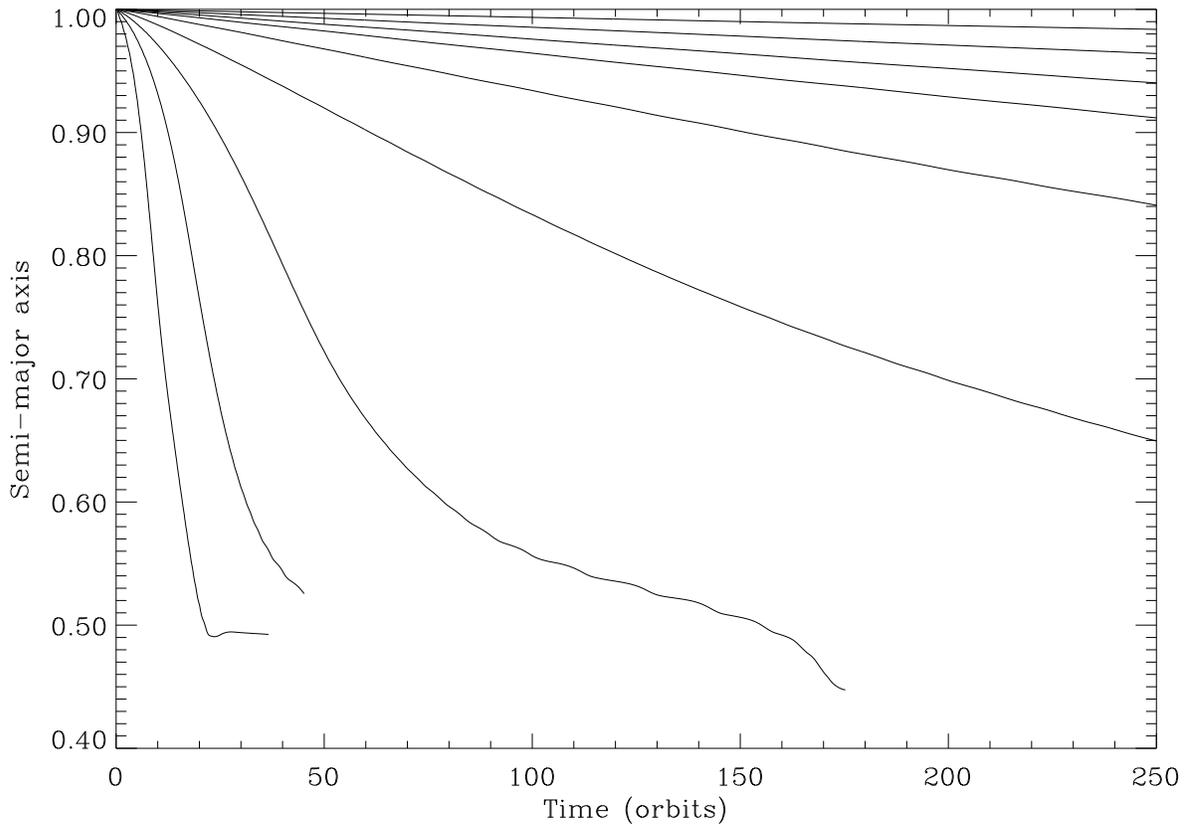}
  \caption{Semi-major axis as a function of time, for the different values of $S_n$, 
    $n$ ranging from $0$ to $8$ from top to bottom. The behavior is
    meaningless when $a$ gets close to the grid inner boundary,
    located at $R_{\rm min}=0.4$.}
  \label{fig:a-vs-t}
\end{figure}

\begin{figure}
  \plotone{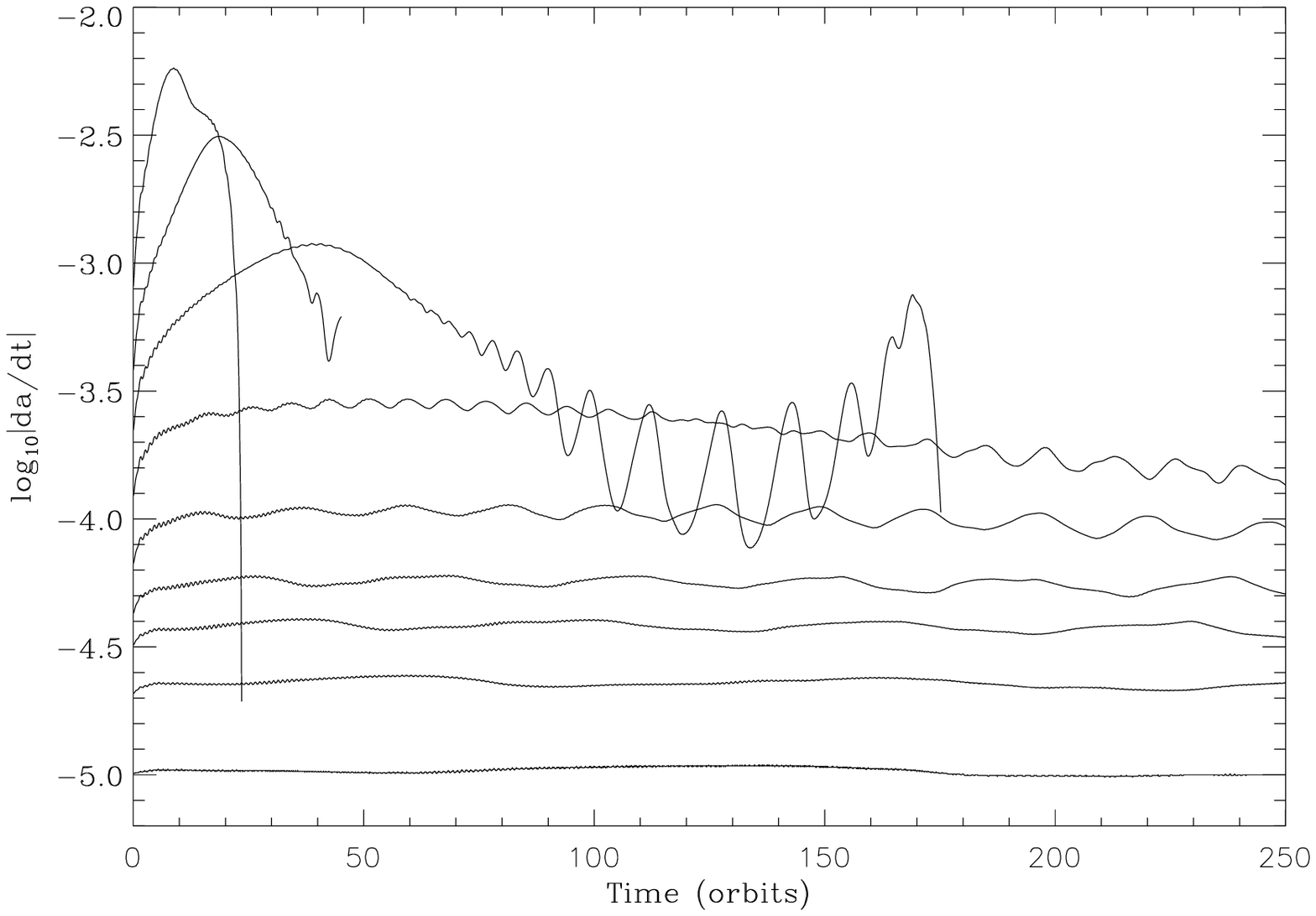}
  \caption{Drift rate as a function of time. 
$n$ ranges from $0$ to $8$ from bottom to top.
  \label{fig:da-vs-t}}
\end{figure}

\begin{figure}
  \plotone{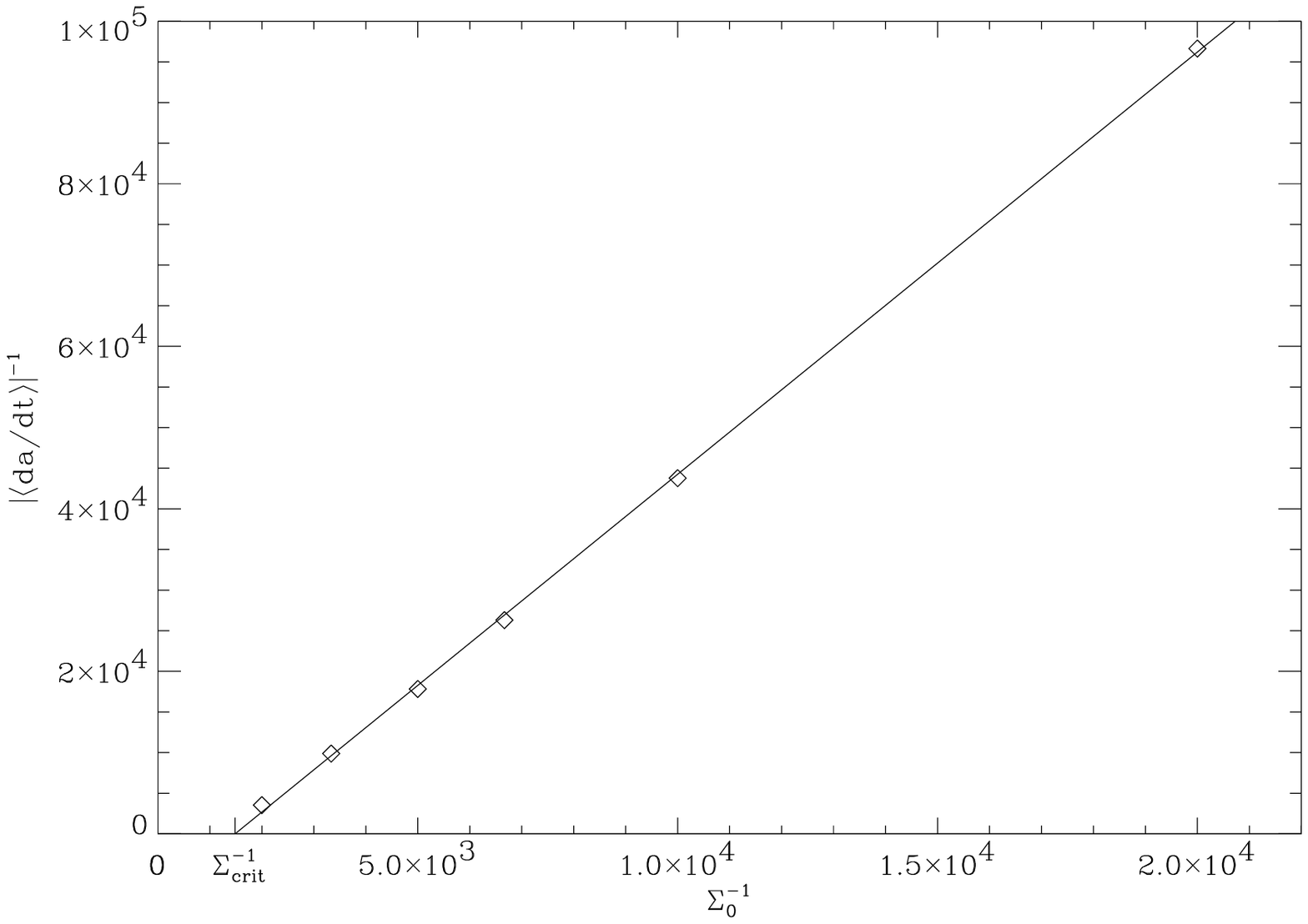}
  \caption{Inverse of average drift rate as a function of the inverse of disk surface density.
  \label{fig:rel}}
\end{figure}

We show the temporal behavior of the planet semi-major axis for our
nine runs at~Fig~\ref{fig:a-vs-t}. The origin of time is chosen at the
planet release. The first $477$~orbits are therefore not shown.
One can already notice that the radial drift rate is not proportional
to the disk mass. The heaviest disk is indeed ``only'' $40$~times
heavier than the lightest one we consider. For the lightest disk the
planet migrates about $\sim 1.5$~\% inwards over $250$~orbits, whereas
for the heaviest disk the planet is already at $r=0.5$ after about
$20$~orbits.  Fig.~\ref{fig:da-vs-t} shows the migration rate as a
function of time. We see that it is relatively constant for the runs
$S_0$ to $S_4$, that a slight global variation can be seen for the run
$S_5$, and that the other runs, $S_6$ to $S_8$, display a strongly
variable migration rate, which peaks at very large values. Some
oscillations in the migration rate for $S_0$ to $S_6$ can be seen.
These can be identified with the planet crossing the boundaries of
mesh zones (i.\ e. the period for these oscillations is $|\Delta
r/\dot a|$, where $\Delta r$ is the radial zone size, which explains
why these oscillations are slower for the smaller migration rates).
This also gives us an idea of the accuracy of the numerical scheme and
of the torque dependency on the planet placement with respect to a
mesh zone. This accuracy is satisfactory except for the run $S_6$
which exhibits large amplitude variations, but these are likely not
relevant since the planet is then close to the grid inner boundary.
We can check that the drift rate grows faster than linearly with the
disk surface density for runs~$S_0$ to $S_5$. We estimate the average
migration rate for these runs, over the whole time interval for runs
$S_0$ to $S_4$, and over the time interval $40$ to $80$~orbits for the
run $S_5$ (we discuss this choice later). Fig.~\ref{fig:rel} shows
$|\dot a|^{-1}$ as a function of $\Sigma^{-1}$.
The points are relatively well aligned, as was expected from
Eq.~(\ref{eq:check}). A linear regression fit allows one to determine
the critical surface density for runaway, and yields $S_{\rm
  crit}=6.7$.  The runs $S_6$ to $S_8$, which correspond to disk
surface densities larger than this runaway threshold, do indeed
exhibit a very fast migration and a strongly time variable migration
rate.  One can also understand the time behavior of the drift rate in
the first orbits of Fig.~\ref{fig:da-vs-t}.  As can be seen in
Eq.~(\ref{eq:estab}), the timescale $\tau_d$, over which the migration
rate tends to its limiting value given by~Eq.~(\ref{eq:ode1}),
increases when the disk surface density approaches its critical value.
This is precisely the trend that we see in our runs. This is the
reason we took $40$ orbits as a lower time value for estimating the
average migration rate for run $S_5$. The higher time value of
$80$~orbits comes from the fact that by then the planet has already
migrated a sizable fraction of its initial distance to the star such
that its coorbital mass deficit may have been significantly altered.

\begin{figure}
  \plotone{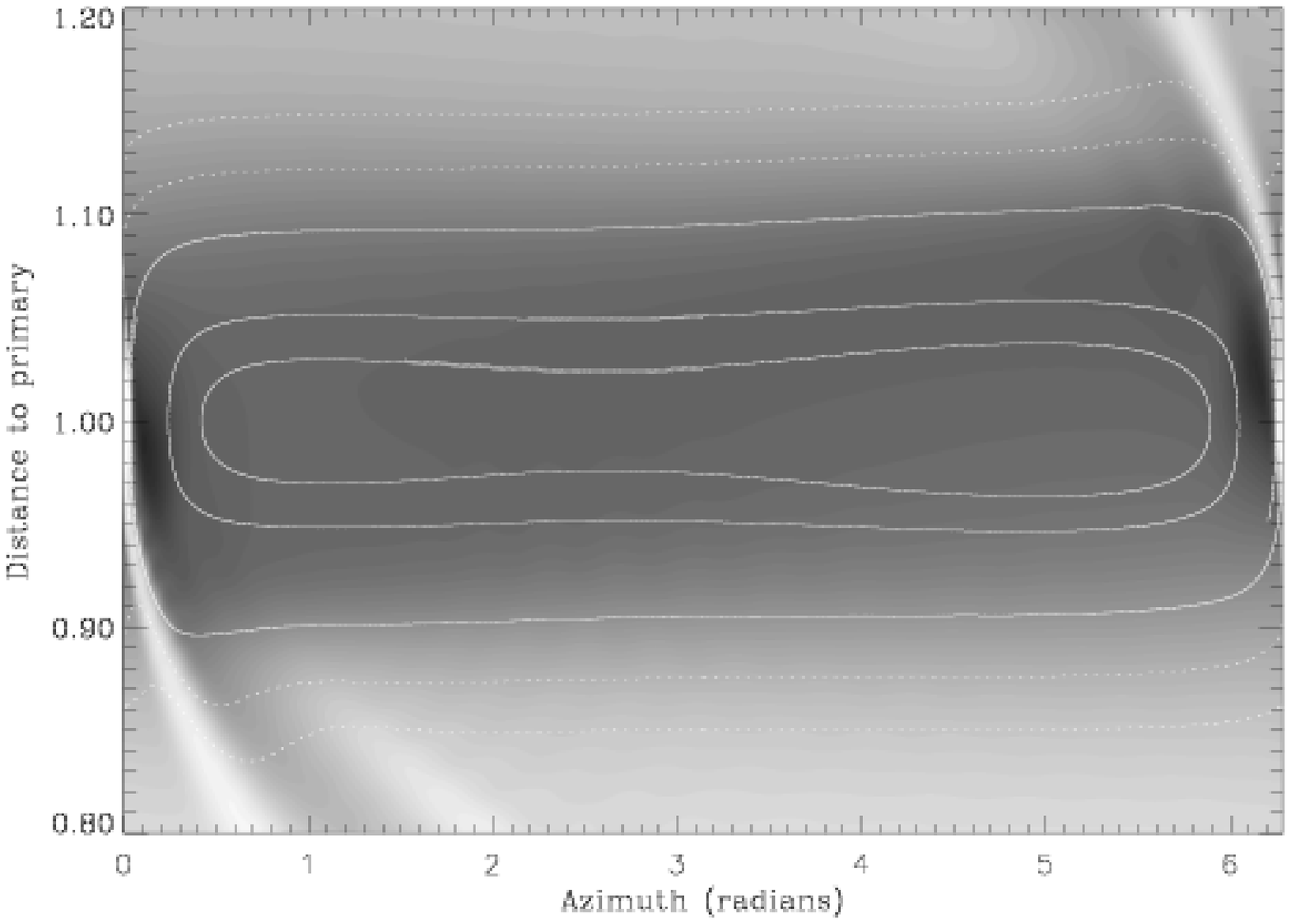}
  \caption{Surface density and streamline aspect for any run $S_0$ to $S_8$ just before the planet is
    released. The circulating streamlines are dotted, while the
    librating ones are solid.
  \label{fig:sl}}
\end{figure}

\begin{figure}
  \plotone{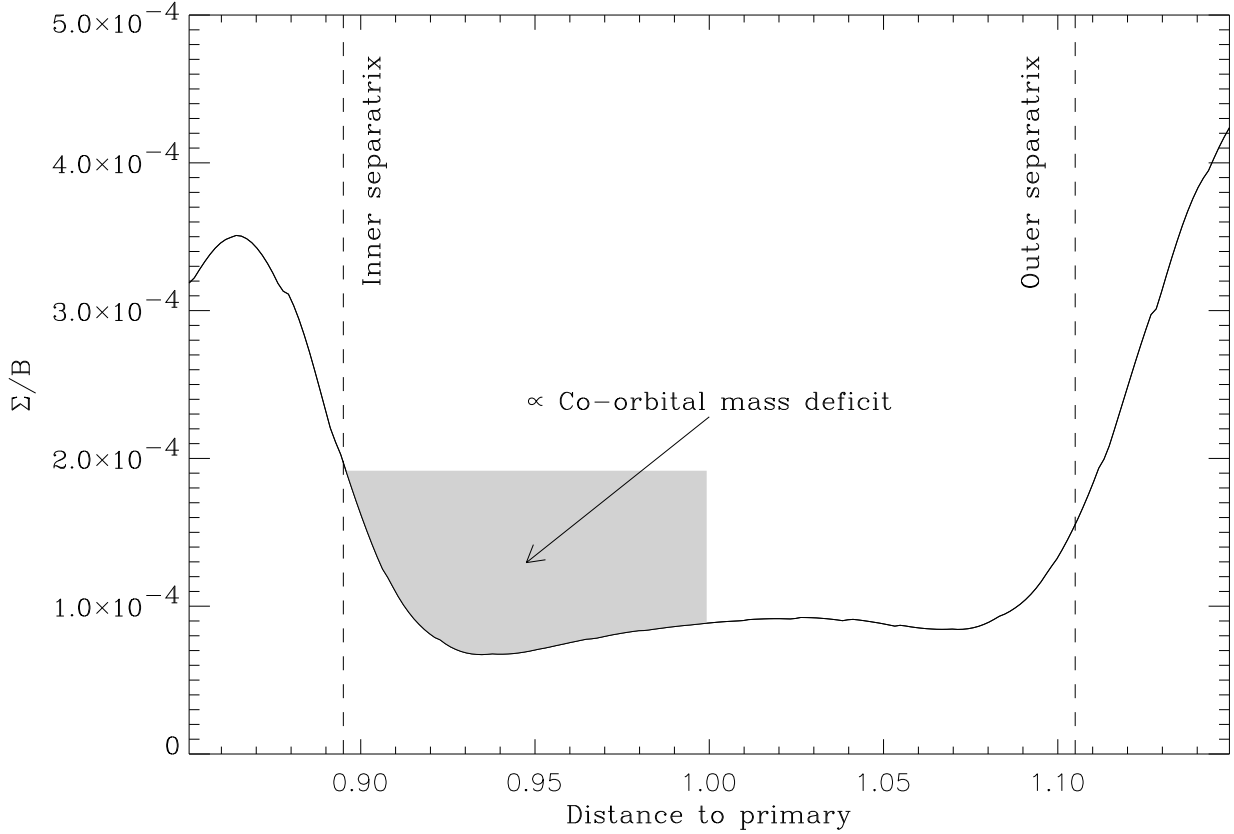}
  \caption{Coorbital mass deficit estimate from run~$S_1$.
    As can be seen the profile is depleted in the coorbital region.
    The coorbital mass deficit is related to the shaded area.  The
    profile is relatively symmetric inside of the coorbital region,
    which indicates that the corotation torque is saturated (cancels
    out) in the absence of migration.  The inner separatrix position
    is $x_s=-0.105$. The coorbital mass deficit, as shown here, has to
    be evaluated from the depleted profile shape between the orbit and
    the upstream separatrix.  For the case of an outwards migration
    the right part of the profile should have been used. As we
    consider in this case a $\Sigma \propto r^{-3/2}$ profile, this
    would make no difference as the unperturbed specific vorticity
    profile is flat.
  \label{fig:cmd}}
\end{figure}

We finally check whether the critical surface density corresponds to a
coorbital mass deficit comparable to the planet mass. We display at
Fig.~\ref{fig:sl} the surface density in the coorbital region (in a
$\theta,r$-plane) and we superimpose a few streamlines.
This allows us to get an estimate of the position of the separatrices,
which we find at $x_s\approx \pm 0.1a_0$.  Once one knows the location
of the separatrices, one can estimate the coorbital mass deficit.
Fig.~\ref{fig:cmd} shows the inverse specific vorticity profile for
run~$S_1$, from which one can estimate a coorbital mass deficit
$\delta m=3.44\cdot 10^{-5}$. This latter needs therefore to be
$M_p/\delta m=8.7$ times more massive to fulfill the runaway
condition, from which we conclude that the runaway should occur for
$S=8.7$. This value is $30$~\% larger than the value inferred from the
linear regression fit.  The agreement is not surprisingly rough,
probably owing to the various assumptions made to derive
Eq.~(\ref{eq:cmd}).

\subsection{Additional runs}
In addition to the main runs involving freely migrating planets, we
have performed a series of additional runs in which the migration rate
$\dot a$ is fixed in order to check the behavior described by
Eq.~(\ref{eq:tqexp}). Namely, in these runs, the planet was held on a
fixed $\dot a$ trajectory for 200 orbits.  The run was started with
semi major axis $a_0=1-150\cdot 2\pi \dot a$ and ended with semi major
axis $a_1=1+50\cdot 2\pi\dot a$, so that in all the runs the planet
has semi major axis $a=1$ at $t=150$~orbits.  The planet is given an
instantaneous orbital frequency equal to $\Omega_K(a)$. The runs are
performed as before in the frame corotating with the guiding center.
The resolution and other numerical values are unchanged with respect
to what is described in the previous section.  The values adopted for
$\dot a$ are $\dot a= D\times 5\cdot 10^{-5}$, where $D$ is an integer
ranging from $-3$ to $+6$. The surface density in all these runs is
$\Sigma_0=10^{-4}$, corresponding to $S=1$.

\begin{figure}
  \plotone{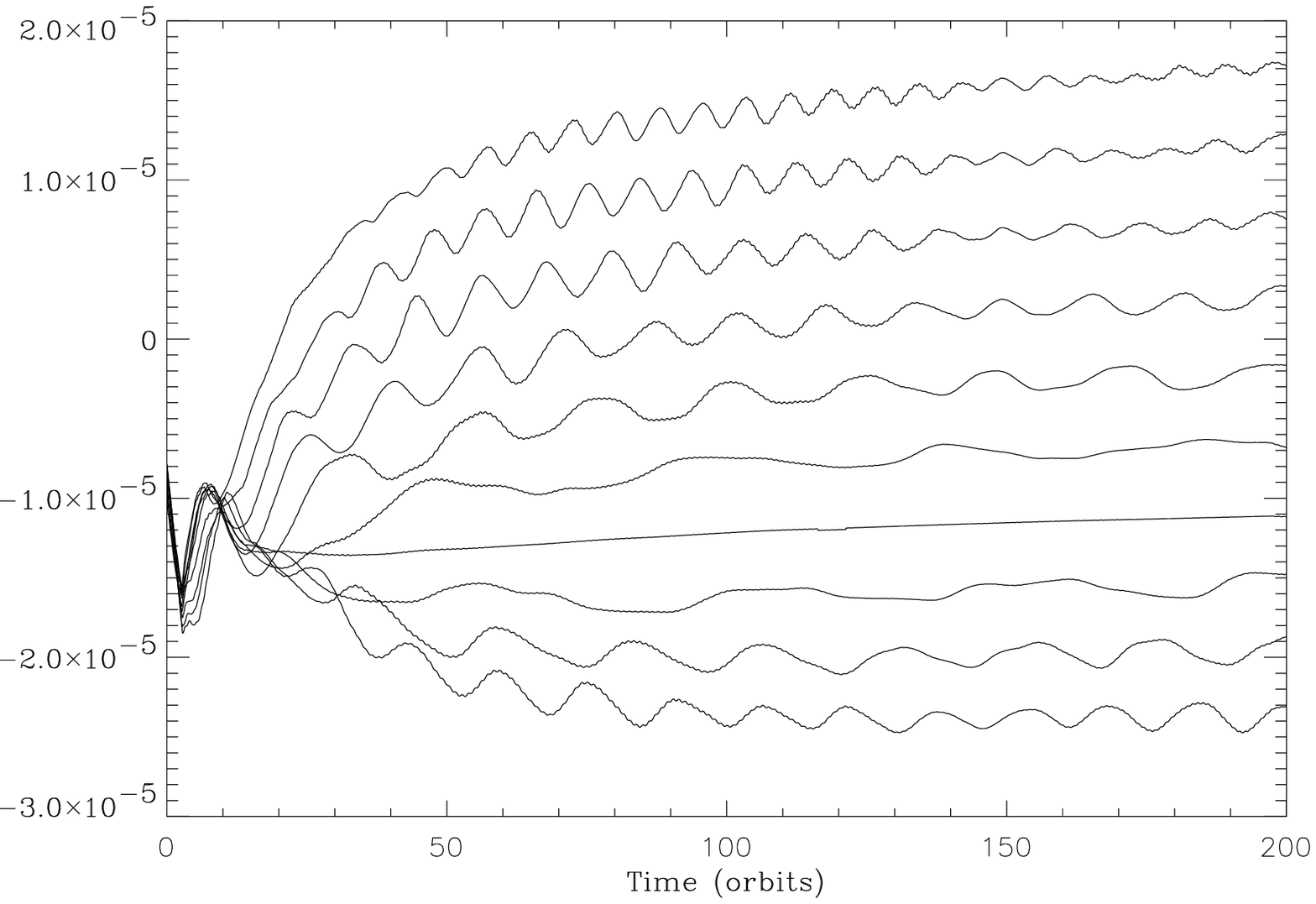}
  \caption{Measured total specific torque value (smoothed over a $5$~orbit temporal window) as a function
    of time, for runs $D=-3$ to $+6$. The value of $D$ increases from
    bottom to top.
\label{fig:gammaadot1}
}
\end{figure}

\begin{figure}
  \plotone{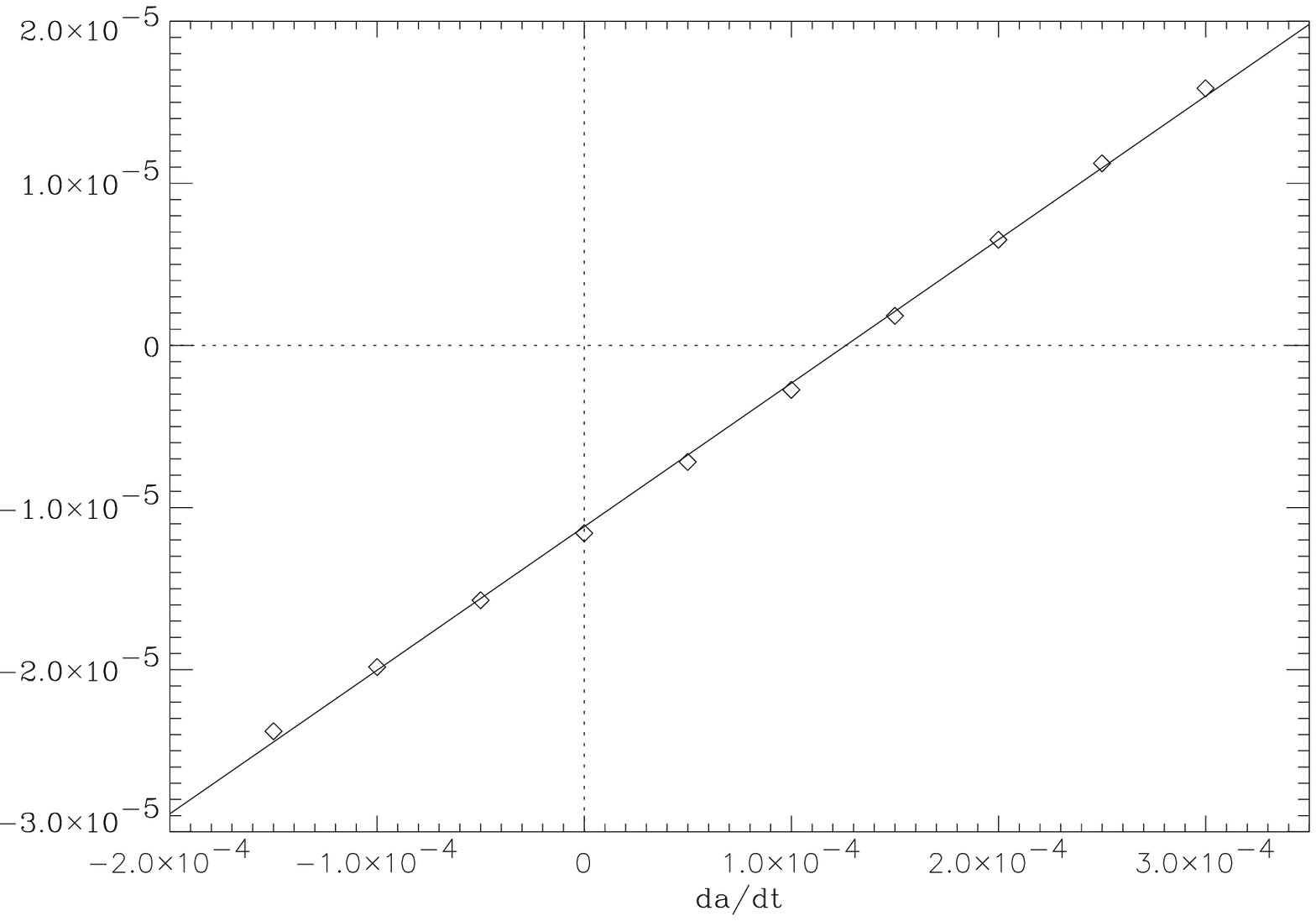} \caption{Average total specific torque value over
  the time interval $[100,200]$~orbits for the runs of
  Fig.~\ref{fig:gammaadot1}, as a function of the imposed drift
  rate. The solid line shows the linear regression fit performed on
  the runs.  The torque value for $da/dt=0$ corresponds to the
  differential Lindblad torque (one can notice that this value is much
  smaller than what is given by a linear estimate, mainly because of
  the strong perturbation of the surface density profile). Also note,
  as stated in \S~\ref{sec:setup}, that the resolution used in our
  runs is just barely sufficient to get a proper estimate of the
  Lindblad torques for the very thin disks that we consider, whereas
  it is largely enough to describe properly coorbital effects.
\label{fig:gvad}
}
\end{figure}

Fig.~\ref{fig:gammaadot1} shows the torque as a function of time for
each of the ten runs performed, and Fig.~\ref{fig:gvad} shows the
average value of the torque as a function of $\dot a$ (the average is
taken over the time interval $[100,200]$~orbits, which eliminates
initial transient behavior occurring over the first $100$~orbits, and
which ensures that the average semi major axis of the planet over this
time interval is the same for all the runs, and is $\bar a=1$.)

The torque modulation as the planet sweeps radially the mesh zones is
again apparent in Fig.~\ref{fig:gammaadot1}.  These plots confirm the
linear dependence of the total torque upon the migration rate. One can
check that the slope of $\Gamma$ as a function of $\dot a$ is
positive, i.\ e. the feed back is positive, and one can infer from the
slope estimate the critical disk surface density for runaway. We note
$\gamma=\Gamma/M_p$ the specific torque acting on the planet. We can
write, since $\ddot a=0$:
\begin{eqnarray}
\gamma_{\rm tot}&=&\gamma+\Delta\gamma_{\rm LR}\nonumber\\
&=&2B_pa\dot a\frac{\delta m}{M_p}+\Delta\gamma_{\rm LR}.
\end{eqnarray}
We also have the relationship:
\begin{equation}
\label{eq:deltamsigma}
\delta m=M_p\Sigma/\Sigma_{\rm crit},
\end{equation}
since $\delta m$ scales with $\Sigma$, and since the runaway starts
for $\delta m=M_p$.  If we call ${\cal S}$ the slope of $\gamma_{\rm
  tot}$ as a function of $\dot a$, then we have:
\begin{equation}
2B_pa\frac{\Sigma}{\Sigma_{\rm crit}}={\cal S}.
\end{equation}
The linear regression fit displayed at Fig.~\ref{fig:gvad} yields
${\cal S}=0.088$, from which one infers $\Sigma_{\rm crit}=5.7\cdot
10^{-4}$, in good agreement (within $\sim15$~\%) with the estimate
given by the linear regression fit of Fig.~\ref{fig:rel}. This is
however only in rough agreement with the result given by the coorbital
mass deficit estimate.
 
These additional runs also enable one to check whether the assumption
that the horseshoe zone width does not depend upon the drift rate is
valid or not. We evaluate the distance $x_s$ of the inner separatrix
to the orbit at time $t=150$~orbits for all the runs.  We find that
$x_s = 0.105\pm0.005$ and that this quantity exhibits no systematic
trend with $\dot a$.

\subsection{Corotation torque in the fast migration regime}
\label{sec:fast}
\begin{figure}
\plotone{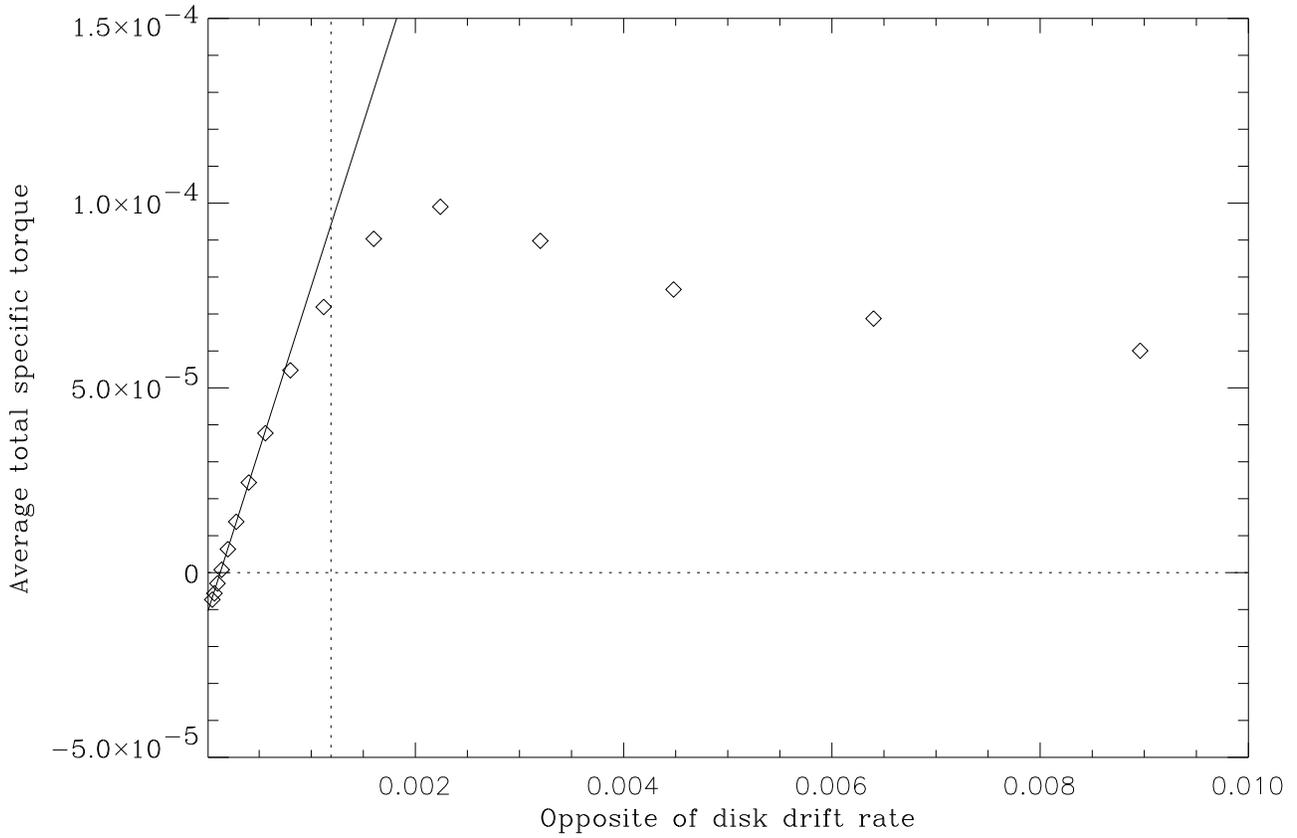}
\caption{Total specific torque acting on the planet as a function of the opposite of the imposed
  disk drift, with a linear scale on the $x$-axis.  The
  solid line shows the linear regression fit that was
  obtained from the data of Fig.~\ref{fig:gvad}. The vertical dotted
  line shows the critical drift rate $|\dot a_c|$ for fast
  migration. 
\label{fig:large_adot}}
\end{figure}

\begin{figure}
\plotone{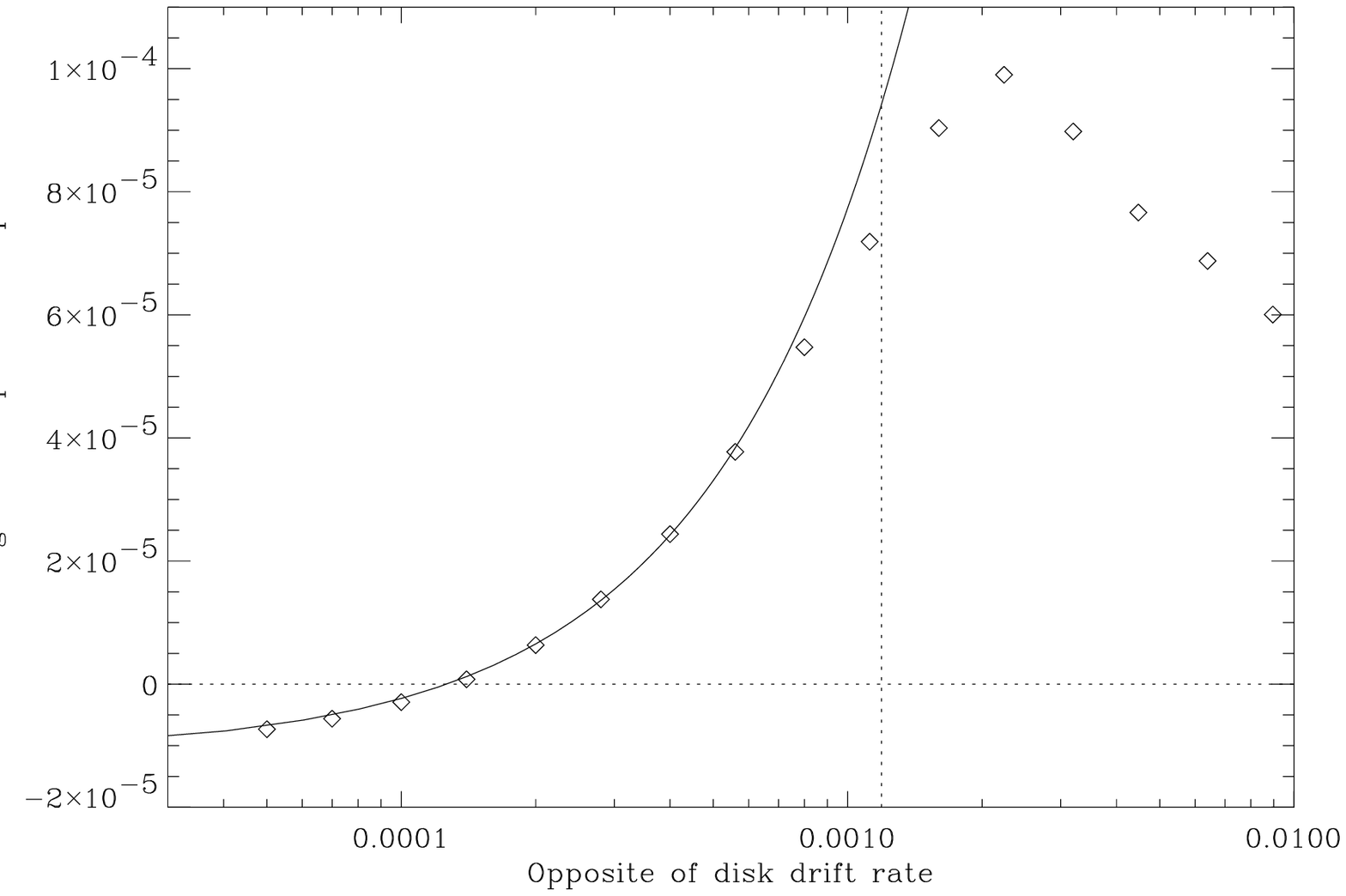}
\caption{Same as Fig.~\ref{fig:large_adot} with a logarithmic scale on the $x$-axis. 
This plot offers a number of similarities with the torque versus
viscosity relationship \citep{m02}.
\label{fig:large_adot2}}
\end{figure}

The previous section illustrates the linear dependency of the
coorbital corotation torque on the drift rate in the slow migration
regime (i.\ e. $\dot a \ll |A_p|x_s^2/\pi a$). In order to investigate
the fast migration regime, and in particular in order to get an idea
of the migration rates that can be achieved in a runaway episode, it
is of interest to know the $\Gamma(\dot a)$ relationship for large
values of $\dot a$.  The problem in that regime is that if an external
operator imposes a fixed large drift rate to the planet, this latter
sweeps a sizable fraction of its initial orbital radius in a very
short time, and no reliable value can be measured for the
corresponding torque. A workaround can be found as follows. As the
torque that we aim at measuring arises from a relative drift of the
disk material with respect to the planet horseshoe region, one can
mimic this drift by adding an external specific torque to the disk
material in order to make the fluid elements drift radially, while the
planet is held on a fixed circular orbit. That way a steady state
situation can be achieved, which allows a precise measurement of the
torque even for very large drift rates. Namely, we performed a number
of runs for which:
\begin{enumerate}
\item the planet is held on a fixed circular orbit with radius $a=1$,
\item the disk material undergoes an additional, external specific torque, with expression:
\begin{equation}
\label{eq:specific_torque}
\gamma_{\rm add} = \frac{2B(r)av_d\Sigma_p}{\Sigma_0(r)},
\end{equation}
where $v_d$ is the disk material radial drift velocity at the planet
orbit and $\Sigma_0(r)$ the unperturbed surface density; this
expression ensures that the radial drift velocity in an axisymmetric
situation, that is $u=v_d\frac{\Sigma_p a}{\Sigma_0(r) r}$,
corresponds to a steady state situation (i.\ e. that $\partial[\Sigma
r u]/\partial r=0$),
\item a source of disk material with the adequate surface density is
  set at the grid outer / inner boundary for $v_d<0$ / $v_d>0$. This
  ensures that no disk depletion occurs at large $v_d$, which would
  modify the surface density profile and therefore would affect the
  torque value.
\end{enumerate}
This torque prescription leads to the radial drift of any structure in
the surface density profile in much the same way as for a viscous
drift, but contrary to this latter it does not lead to a radial spread
of the profiles.  We ran $16$ such configurations, with $v_d^i=-5\cdot
10^5\times 2^{i/2}$ ($0\leq i\leq 15$). As this corresponds to an
inwards drift of the disk material, the upstream separatrix is the
outer separatrix, and therefore the torque value should correspond to
the ones measured at the previous section for positive values $\dot
a$. The results are presented at Figs.~\ref{fig:large_adot}
and~\ref{fig:large_adot2}.
Clearly there is a satisfactory agreement between these results and
the results obtained at the previous section for the case of true slow
migration, as can be seen on the left part of the plots.  This
validates this method as an alternate way of measuring the corotation
torque dependence on the migration rate.
This agreement can be understood using similar arguments as the ones
used by \citet{m01} to evaluate the corotation torque in a viscous
disk. The librating fluid elements define a trapped region, the
angular momentum of which is therefore constant in time. The external
torque applied on this region is therefore exactly transmitted to the
planet in a steady state situation. One can easily show that to lowest
order in $x/a$ an expression similar to Eq.~(\ref{eq:tqexp_one}) is
obtained for the corotation torque.  The agreement between the
measured torques for a drifting planet and for a backward drifting
disk, as well as the similarity of the flow topology in the
$(\theta,r-a)$ plane in either case, suggests that the torque
measurement for a drifting disk, even in the fast migration limit,
gives a reasonable idea of the behavior of the corotation torque on a
migrating planet. This new method also offers the advantage that the
planet is fixed with respect to the grid, and therefore there is no
torque modulation as observed in Fig.~\ref{fig:gammaadot1} when the
planet sweeps the mesh zones, and thus it enables us to get a more
precise estimate of the disk torque.  The critical drift rate for fast
migration is:
\begin{equation}
|\dot a_c|=\frac{|A_p|x_s^2}{2\pi a}.
\end{equation}
Below this rate, all of the disk fluid elements crossing the upstream
separatrix undergo a horseshoe like close encounter with the planet,
and contribute to the corotation torque, while above this rate, some
of them miss the planet. In the fast regime, which corresponds to the
right part of the plots of Figs.~\ref{fig:large_adot}
and~\ref{fig:large_adot2}, the corotation torque reaches a maximum
value and then slowly decays, while its characteristic order of
magnitude is:
\begin{equation}
\label{eq:tqfast}
\Gamma_{\rm fast} = 2B_pa\,\delta m\, \dot a_c.
\end{equation}
In the series of runs presented in section~\ref{sec:results_run_1},
$\delta m$ scales with the disk surface density. The maximum drift
rate should therefore roughly scale with the disk surface density
during a runaway episode. This is in agreement with the results
displayed at Fig.~\ref{fig:da-vs-t}.

\subsection{Outward runaway migration}
Since in the runaway regime the differential equation governing the
time evolution of the planet semi-major axis is second order in time,
it is formally possible to have an outward migration for an adequate
choice of the initial $a$ and $\dot a$. For an outward migration, the
upstream separatrix is the outer one. The higher the (inverse of)
specific vorticity jump across that separatrix, the easier it is to
get an outward migration. Therefore, outwards migration should be
easier to get for shallower surface density profiles (corresponding to
steeply increasing profiles of $\Sigma/B$). The weakening of the
differential Lindblad torque for shallower surface density profiles
\citep{tanaka} plays in the same direction, since this torque tends to
favor inward runaway rather than outward runaway. In order to
illustrate this trend, we have performed a series of runs in which we
hold the planet on a fixed $\dot a>0$ orbit for $100$~orbital times,
and then release it (i.\ e. we allow it to freely migrate under the
action of the disk torque).  The planet mass, disk aspect ratio and
viscosity, the grid resolution and the numerical algorithm were
strictly the same as in section~\ref{sec:results_run_1}, the disk
surface density was $\Sigma(r)=10^{-3}r^{\alpha}$, corresponding for
$r=1$ to ${\cal S}=10$, i.\ e.  approximately $50$~\% above the
runaway critical surface density. We tried four values for $\alpha$:
$-3/2$, $-1$, $-1/2$, and $0$, corresponding to an increasingly
shallower surface density profile. The starting semi-major axis is
$a=0.7$, and the semi-major axis at the time of release
($t=100$~orbital times) is $a=1$, which ensures that in the four runs
the disk surface density at the planet orbit is the same at the
release time, and only differs by its slope.
\begin{figure}
  \plotone{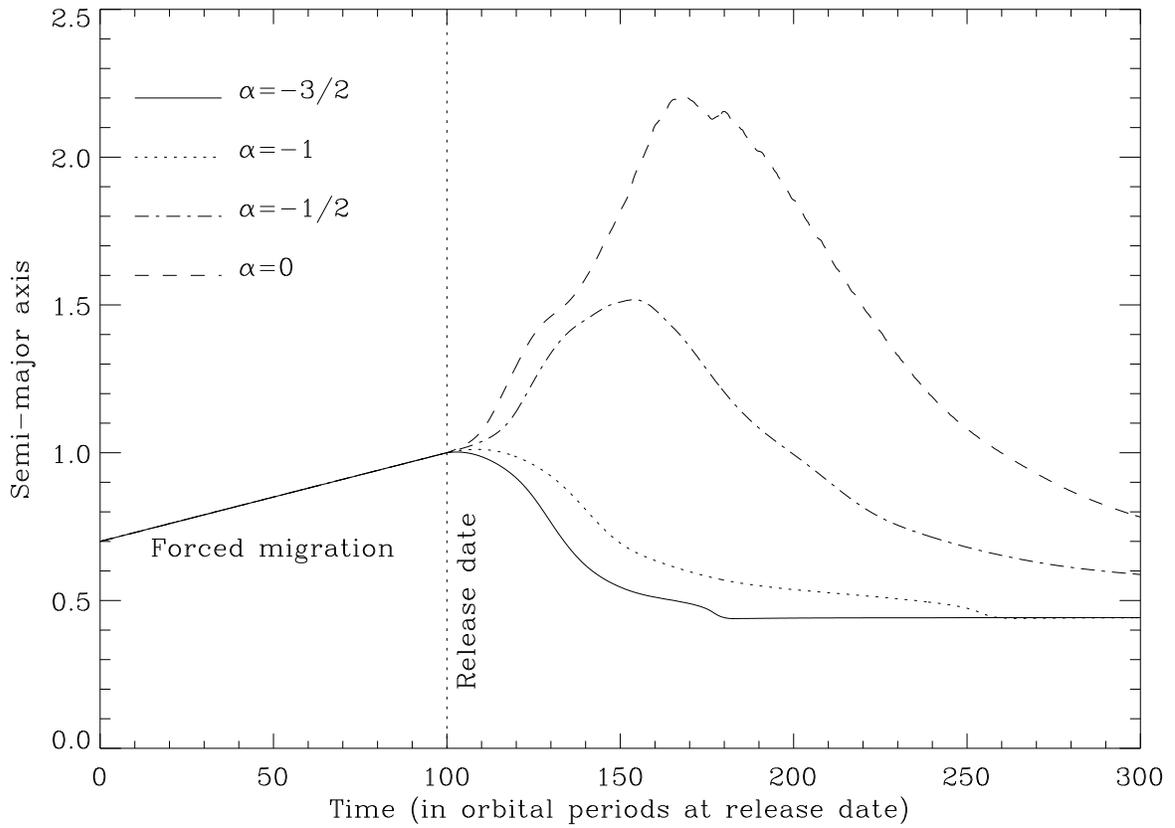}
  \caption{Time evolution of the planet semi-major axis for the four runs detailed in text. An outward
    runaway occurs for the two shallowest surface density profiles.
    The inner grid boundary is located at $r=0.4$, and the curves lose
    significance whenever the planet get close to this inner boundary.
\label{fig:outwards}
}
\end{figure}
We see in Fig.~\ref{fig:outwards} that the planet undergoes an inward
runaway for the two steepest surface density profiles, and an outwards
runaway for the two shallowest. The maximum $\dot a$ is of the same
order of magnitude in the four cases, and the runaway starting time,
corresponding to the short exponential regime in which migration can
be considered as slow ($|\dot a|<|\dot a_c|$), is of the order of ten
orbits, comparable to the outermost horseshoe libration time. This
illustrates that a common mechanism is at work for the inward and
outward runaways, and that these are tightly linked to the coorbital
dynamics.

\section{Discussion}
\label{sec:discussion}
\subsection{Occurrence of runaway migration}

\begin{figure}
  \plotone{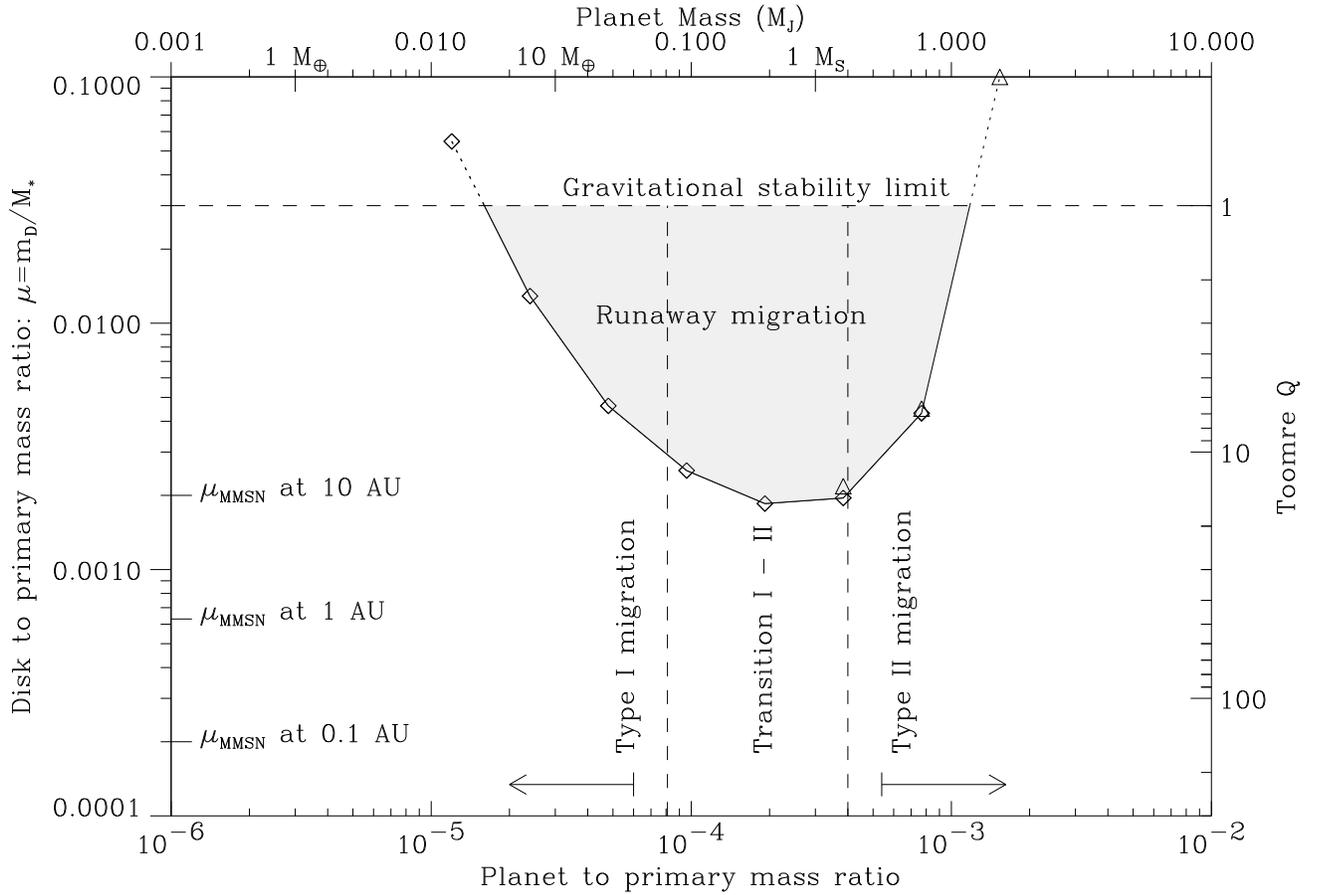}
  \caption{Runaway migration domain for a 3~\% aspect ratio disk with viscosity $\nu=10^{-5}$. The
    domain horizontal upper limit corresponds to the $Q=1$
    gravitational stability limit (which translates into $\mu=h$). The
    left vertical dashed line shows the limit of linear type~I
    migration, and corresponds to $R_H=H$, where $R_H$ is the planet
    Hill radius. Note that \citet{miyo99} find a threshold for
    non-linear effects for even lower masses ($R_H=H/2$), which
    corresponds to a factor~$8$ in the mass. The right vertical dashed
    line corresponds to the viscous gap opening criterion $q>40/{\cal
      R}$, where ${\cal R}=a^2\Omega(a)/\nu$ is the disk Reynolds
    number at the planet orbit \citep{p84}, beyond which the disk is
    split into an outer and an inner disk, and the planet is locked in
    the disk viscous drift (type~II migration). Diamonds indicate the
    critical values found using method~1 (see text), while triangles
    indicate critical values found using method~2. The tick marks on
    the left axis represent the reduced disk mass of the minimum mass
    solar nebula respectively at $10$, $1$ and $0.1$~AU. The upper
    axis shows the planet mass in Jupiter masses if one assumes that
    the central object has one solar mass.
\label{fig:RWlim1}
}
\end{figure}

\begin{figure}
  \plotone{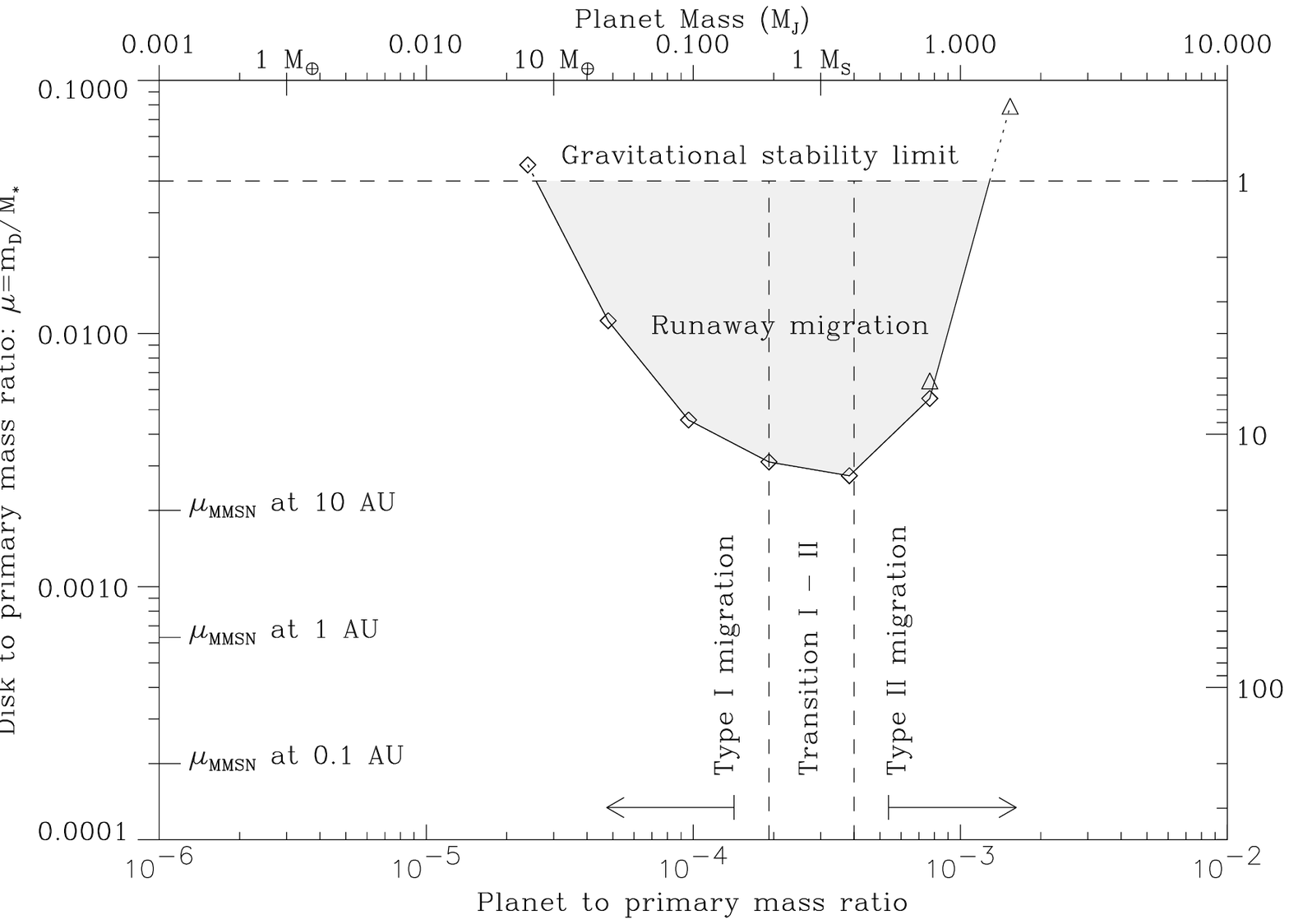}
  \caption{Same as Fig.~\ref{fig:RWlim1} for a $h=4$~\% aspect ratio disk.
\label{fig:RWlim2}
}
\end{figure}

\begin{figure}
  \plotone{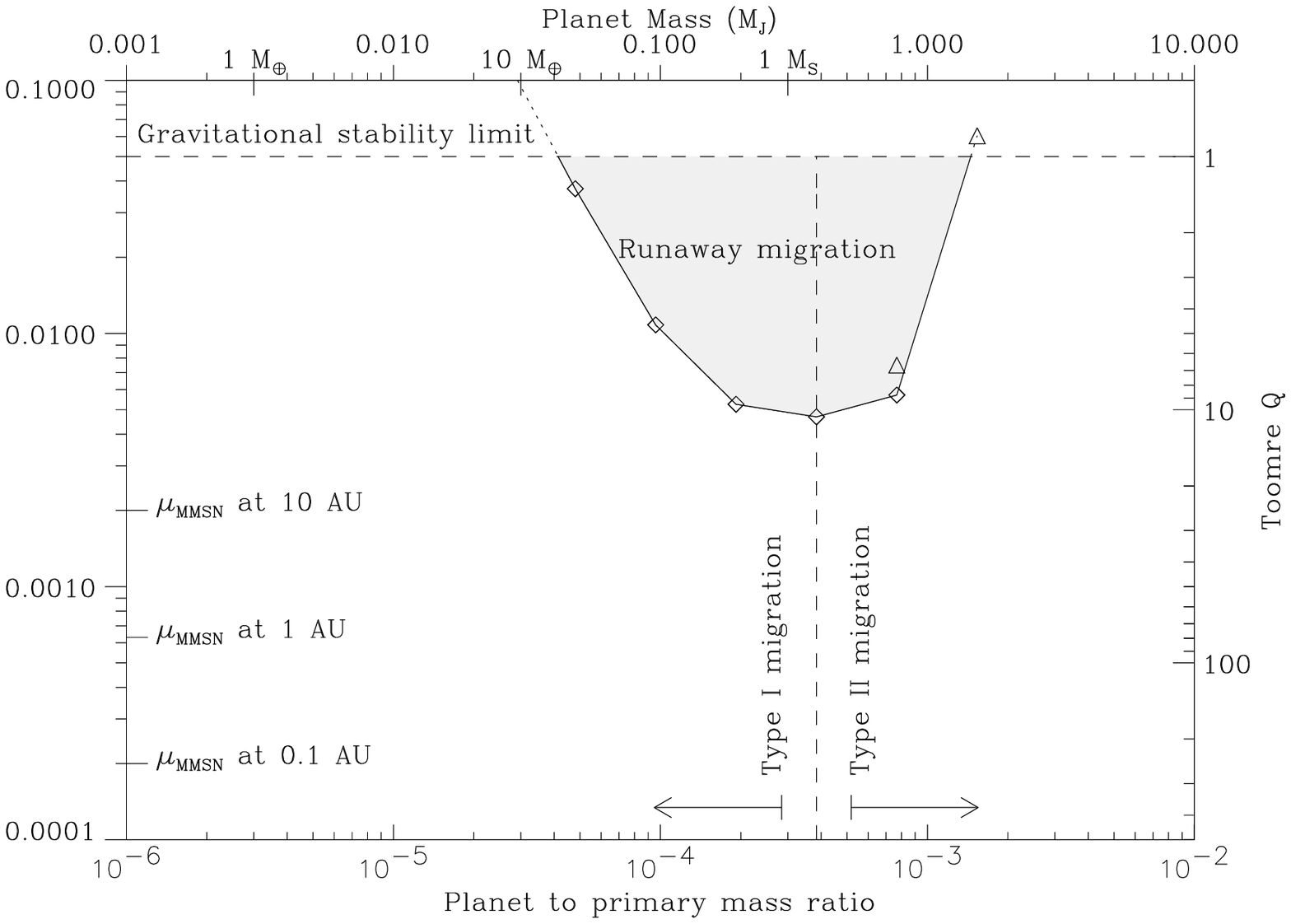}
  \caption{Same as Fig.~\ref{fig:RWlim1} for a $h=5$~\% aspect ratio disk
\label{fig:RWlim3}
}
\end{figure}

In order to assess the importance of runaway migration in
protoplanetary nebulae, we have tried to delineate the runaway
migration domain borders in a (planet mass, disk mass) space, while
keeping the disk aspect ratio and viscosity fixed.  All our disks have
$\Sigma_0(r) \propto r^{-3/2}$, and a viscosity $\nu=10^{-5}$. We
tried three values of the disk aspect ratio: $h=0.03$, $0.04$ and
$0.05$, corresponding respectively to values for the $\alpha$
parameter $1.1\cdot 10^{-2}$, $6.3\cdot 10^{-3}$ and $4\cdot 10^{-3}$.
We call disk mass the quantity $m_D=\pi a^2\Sigma_0(a)$, and disk
reduced mass the quantity $\mu=m_D/M_*$.  The critical disk mass for
runaway depends on the planet mass, as it depends on the dip shape
around the orbit and on the position of the separatrices. In order to
determine the disk critical mass, we measure the disk torque exerted
on the planet, which is held on a fixed circular orbit, for the case
where we exert no additional torque on the disk material (we do not
impose any additional disk radial drift other than the one arising
from its viscous evolution), and for the case where we exert an
additional torque on the disk material (which corresponds to imposing
an additional radial drift of the disk material with velocity $v_d$).
In the first case we measure a torque $\Gamma$, and in the second case
a torque $\Gamma'$.  From Eq.~(\ref{eq:tqexp}) and
section~\ref{sec:fast}, these two torques can be written respectively
as:
\begin{eqnarray}
\Gamma&=&\Delta\Gamma_{\rm LR}\nonumber\\
\Gamma'&=&\Delta\Gamma_{\rm LR}-2B_pa v_d\,\delta m,
\end{eqnarray}
where $\Delta\Gamma_{\rm LR}$, the differential Lindblad torque, is
assumed to be independent of the disk radial drift velocity, and where
the ``static'' part of the corotation torque vanishes since we
consider initially a uniform specific vorticity disk.  As the planet
is held on a fixed circular orbit and a stationary state is reached,
one can use Eq.~(\ref{eq:deltamsigma}), which yields:
\begin{equation}
\label{eq:sigmacrit}
\Sigma_{\rm crit}=-\frac{2B_pM_pv_d\Sigma}{\Gamma'-\Gamma}=-\frac{2B_pv_d\Sigma}{\gamma'-\gamma},
\end{equation}
where $\gamma=\Gamma/M_p$ and $\gamma'=\Gamma'/M_p$.  The value $v_d$
must be chosen small enough so that it corresponds to the slow
migration limit, and large enough that it allows an accurate
measurement of $\gamma'-\gamma$.  Preliminary runs have shown that an
accurate estimate of the critical surface density can be reached with
a value of $v_d$ as small as $2\cdot 10^{-5}$, corresponding to a disk
radial drift which amounts to less than one zone radial width over the
whole simulation, which illustrates the fact that even a small
resolution grid, with a reduced number of zones across the horseshoe
region, captures remarkably well most of the features of the
corotation torque, as noted by \citet{m02}.  The measurement of
$\gamma$ and $\gamma'$ can be performed in two different ways:
\begin{itemize}
\item either we perform two different runs with 
constant values of the disk drift, $0$ and $v_d$ (method~1), 
\item or we perform one run with a vanishing additional disk drift,
  and once a steady state is reached we switch the drift to $v_d$.
  The new torque value can then be measured after a horseshoe
  libration time (method~2).
\end{itemize} 
The first method is better suited to small mass planets, for which the
libration time is prohibitively long, while the second method is well
suited for higher mass planets, since it allows almost a 50~\% saving
of CPU time compared to the first method. For intermediate, Saturn
mass planets we used both methods to check that they give comparable
results.  The results are presented at Figs.~\ref{fig:RWlim1},
\ref{fig:RWlim2} and~\ref{fig:RWlim3}.
These plots lead to a number of comments:
\begin{itemize}
\item The thinner the disk, the larger the runaway domain. This
  corresponds to expectations: for a given planet mass and disk
  viscosity (such that the planet mass is smaller than the viscous gap
  opening criterion), the thinner the disk, the deeper the dip opened
  around the orbit, and therefore the larger the coorbital mass
  deficit.
\item In the three cases, the mass most favorable to runaway
  (corresponding to the minimum of the critical disk mass curve) is
  $\sim 0.3-0.4$~$M_J$, or typically a Saturn mass.
\item Runaway migration can be found in relatively massive
  protoplanetary disks (a few times more massive than the minimum mass
  solar nebula, depending on the protogiant semi-major axis).
\item Runaway migration should be common, in such disks, for giant protoplanets which
reach a sizable fraction of a Saturn mass.
\item The right part of the domain boundary shows a steep rise around
  $1$~$M_J$ in the three cases.
\item Even in a disk with $\Sigma\propto r^{-3/2}$ such as the MMSN,
  runaway migration is more difficult as one gets close to the star.
  The coorbital mass deficit scales indeed as $a^2\Sigma(a)\propto
  \sqrt{a}$.
\item The analysis performed breaks down for lower mass planets
  ($R_H<H$), as the torque cannot reliably be estimated through a 2D
  calculation. The runaway limit for the low mass planets at the left
  of the vertical dashed line is likely higher than found in our
  analysis, as in their case only a fraction of the disk vertical
  extent is involved in the coorbital dynamics.
\end{itemize}

\subsection{Additional effects}
The viscosity chosen for these runs is high enough that the dip
viscous time $\tau\sim w^2/3\nu$ is comparable to the libration time
($w \simeq x_s$ being the dip half-width).  This time is also the dip
opening time when the planet is ``switched on'' in the disk, or also
the minimum time that the planet needs to sweep radially its own dip
radial width in order for the surface density profile depression to
follow the planet migration. It is therefore possible to talk
unambiguously of {\em the} disk runaway critical mass for a given
planet, independently of its ``preparation state''. In other words,
whether we release {\em ab initio} the planet in an unperturbed disk,
or whether we hold it on a fixed circular orbit for a few hundreds of
orbits in order to allow it to open a dip before being released makes
no difference: a runaway is observed above the same disk mass, and the
maximum drift rate measured is the same in both cases.
If one tries to delineate the runaway domain for a disk with a much
smaller viscosity, one finds that the estimate strongly depends on the
disk ``preparation state'', contrary to the case of
Figs.~\ref{fig:RWlim1} to~\ref{fig:RWlim3}.  One can achieve indeed a
significant coorbital mass deficit if one holds the planet on a fixed
circular orbit for a sufficient amount of time. Simulations performed
for Saturn mass planets in massive, low viscosity disks display a very
erratic behavior, with an alternation of brief runaway episodes
followed by moderately eccentric ($e\sim 0.01$), halted migration
episodes, the overall drift rate being a relatively small but sizable
($\sim 20$~\%) fraction of the corresponding type~I drift rate.

\begin{figure}
\centering
\plotone{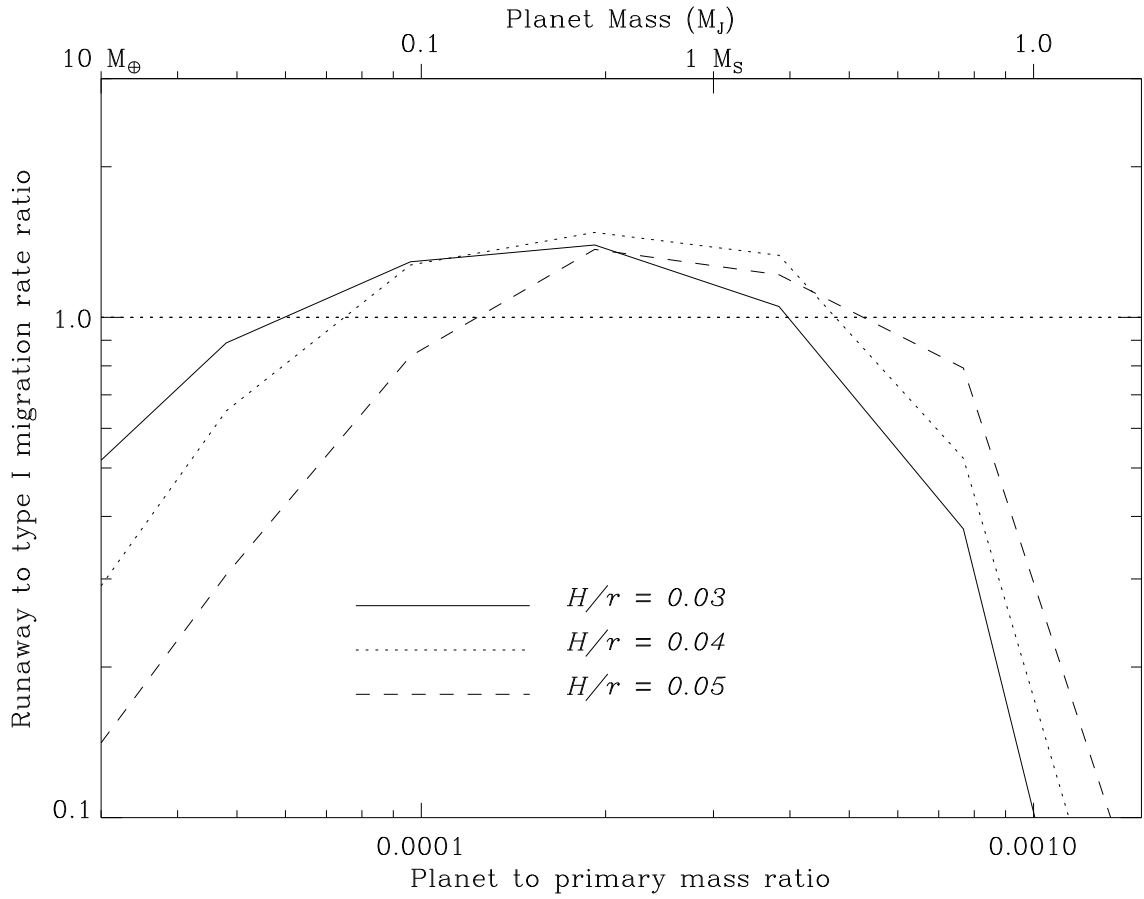}
\caption{\label{fig:RW1ratio}Ratio of runaway to type~I drift rate, at the critical disk mass.}
\end{figure}

It is of interest to evaluate the type~I to runaway drift rate ratio
at the critical disk mass $\mu_c$ for runaway, as a function of planet
mass. At this boundary, Eq.~(\ref{eq:tqfast}) leads to $\dot a=\dot
a_c$ (since $\delta m = M_p$), whereas the type~I drift rate in the
same disk, according to Eq.~(70) of \citet{tanaka}, is given by:
\begin{equation}
\dot a_I=1.38\mu_c qh^{-2}\Omega_pa.
\end{equation}
Fig.~\ref{fig:RW1ratio} shows the ratio of the runaway to type~I
migration rate estimate, as a function of planet mass, for the three
aspect ratios presented in Figs.~\ref{fig:RWlim1}
to~\ref{fig:RWlim3}.  These ratios are estimated at the critical disk
mass for runaway. Since the maximum runaway drift rate scales roughly
with the disk mass (see sect.~\ref{sec:fast}), this ratio should not
vary significantly if one considers disk masses higher than the
critical one.
One can see that this ratio is marginally larger than unity for
slightly sub-Saturn mass planets, and drops considerably for smaller
and larger masses. The runaway drift rate estimate is however still
considerably larger than the type~II viscous drift rate estimate for a
Jupiter mass planet, as it corresponds typically to the third of the
type~I drift rate of a Saturn mass planet.

The main source of the effect we have presented is the drift across
the separatrices of the coorbital region of inner / outer disk
material, in the cases of inward / outward planet drift respectively.
We have assumed that the drift velocity of the planet with respect to
the disk is $\dot a$, thereby neglecting the disk viscous drift rate
$\sim -(3/2)\nu/r$. Taking this additional drift into account would
add a positive contribution to the corotation torque $\sim
3B_p\,\delta m\, \nu$. For the case of a weak coorbital depletion,
$\delta m\propto \Gamma_{\rm LR}/\nu$, so that this additional term
scales as $\Gamma_{\rm LR}$, the one-sided Lindblad torque. This
additional term, coupled to the one-sided Lindblad torque, has been
evaluated to order of magnitude by \citet{m01,m02}, and found to be
small. Since furthermore it does not participate in the feed back and
therefore in the runaway, it is legitimate to have neglected it in the
present analysis.

The ability of the planet to maintain its coorbital mass deficit
during a runaway episode deserves further investigation. Before
runaway, the coorbital region is partially cleared of disk material
under the action of the Lindblad torques. During the runaway drift,
the Lindblad torques are assisted in maintaining the coorbital mass
deficit by the horseshoe dynamics, which traps the coorbital material.
The streamlines are not exactly closed in the $(\theta,r-a)$ space,
however, and the coorbital material can be lost. In that case the
coorbital region is no longer depleted, and the planet switches to
type~I migration, which endows it with a comparable drift rate. No
case has been found in which the runaway is maintained ``forever''. A
single runaway episode however can sizably affect the planet
semi-major axis, by a factor~$2$ or more.

In this work we have neglected accretion onto the protoplanets,
whereas the mass range for which the runaway mechanism is relevant
seems to imply gas accretion. A runaway episode is likely to be
accompanied by a simultaneous accretion of disk material. A simple way
to take into account gas accretion is to consider that a fraction
$\varepsilon$ of the material which enters the coorbital zone through
one separatrix and flows out along the other one, is kept by the
planet. In the absence of accretion, this disk material contribution
to the corotation torque amounts to $\Gamma_2= 4B_pax_s\cdot 2\pi
a\dot a\Sigma(x_s)$. When accretion is considered, the contribution to
the corotation torque reduces to:
\begin{equation}
\Gamma_2'=(1-\varepsilon)\Gamma_2+\frac 12\varepsilon\Gamma_2,
\end{equation}
since the material that is accreted onto the planet loses/gains only
half of the specific angular momentum that it would lose/gain
otherwise. Accretion therefore plays the role of a runaway moderator.
It would be of interest to investigate in detail the interplay between
runaway and accretion, taking consistently into account the corotation
torque feed back reduction and the growth of the horseshoe region.

\section{Summary and conclusions}
\label{sec:summary}

We have evaluated the torque exerted on a protoplanet embedded in a
gaseous disk produced by the fluid elements as they perform a
horseshoe U-turn in the planet vicinity. We have interpreted this
torque as the coorbital corotation torque. This torque exhibits a
dependency upon the planet radial drift rate in the disk which is tied
to the depletion of the coorbital region of the planet, that is to say
to the existence of a dip or gap cleared around the orbit. The sign of
the corotation torque is the same as that of the drift rate.  Hence it
exerts a positive feed back on the migration process.  This feed back
leads to a migration runaway in massive disks when the coorbital mass
deficit is larger than the planet mass. We have checked and
illustrated the main properties of the corotation torque through
customized numerical simulations. These showed the link between
runaway and the coorbital mass deficit. They indicated that the
migration rate for disks with sub-critical masses grows faster than
linearly with disk mass, and they illustrated that the planet drift
rate characteristic response time is the horseshoe libration time.

The occurrence of inward runaway in disks with $\Sigma \propto
r^{-3/2}$ has been investigated, and found to be likely in thin ($h
\leq 5$~\%), massive disks with a mass several times that of the MMSN.
Typically it occurs for Saturn-sized giant protoplanets, but it can
involve planets up to one Jupiter mass for sufficiently massive disks.
The runaway drift rate is found to be comparable to the type~I
estimate given by the differential Lindblad torque that one would
obtain using linear theory even though that would be invalid for a
planet held in a fixed circular orbit 
(because the planet opens a significant dip around its orbit and the
actual differential torque acting on it amounts to a small fraction of
its linear estimate). It turns out that the corotation torque is at
most equal to the differential Lindblad torque in a $\Sigma \propto
r^{-3/2}$ disk. This prevents the possibility of outward runaway in
these disks, and we indeed found no occurrence of outward runaways in
such disks. Additional runs with shallower disk profiles however have
shown that outward runaway can occur, provided the planet is endowed
with an adequate value for $a$ and $\dot a$ (this latter needs to be
established at least over a libration time).

The way we initiated outward runaway was artificial. \citet{ms01} have
exhibited a two planet configuration engaged in an outward migration.
Their outer planet is of Saturn mass, and is therefore a good
candidate for outward runaway in a massive disk. An alternate
possibility to initiate an outward motion is that the disk mass flow
across the orbit is strongly variable in time. If a larger $\dot M$
enters the coorbital zone through the outer separatrix, the corotation
torque may be temporarily large (and positive), providing the seed for
an outward runaway.

When the viscosity is large enough, that the viscous diffusion time
across the horseshoe zone is faster than the time to migrate through
it, relaxation and depletion of the coorbital region occurs in a
sufficiently short amount of time that the coorbital mass deficit and
the disk critical mass for runaway are well defined quantities. The
situation turns out to be much more chaotic in a very low viscosity
disk, although globally the average drift rate of the planet towards
the central object amounts to a sizable fraction of the type~I drift
rate.

Inspection of Figs.~\ref{fig:RWlim1} to~\ref{fig:RWlim3} shows that
in massive disks, sub-Jovian planets can still undergo a type~I like
drift, whereas on previous grounds they were expected to have a much
slower (type~II) drift rate, comparable to the slow viscous disk drift
rate.

The runaway threshold is lower in thinner disks. If protoplanetary
disks are flared (i.\ e.  $\partial \log H/\partial \log r>1$), they
could be extremely thin in their inner regions, which could assist
runaway there.

A runaway episode or a succession of them stops whenever the coorbital
mass deficit that can be achieved is too small. If one assumes a disk
surface density profile $\Sigma(r)\propto r^\alpha$, then the
coorbital mass deficit scales as $\delta m\propto a^2\Sigma(a)\propto
a^{2+\alpha}$. As the runaway condition reads $M_p=\delta m$, one
would get, at the end of the runaway episodes, the relationship
$M_p\propto P^{(4+2\alpha)/3}$, where $P$ is the planet orbital
period.  This indicates a tendency for smaller masses to reach smaller
periods.  A recent analysis by \citet{zm02} indicates a paucity of
planets with masses exceeding a Jupiter mass at small periods~$\sim$~a
few days while sub Jovian mass planets tend to cluster at these small
periods.  The observational data for the larger mass planets is
consistent with a type II migration scenario in which most time is
spent at larger radii \citep{t02}.  However, the distribution of
smaller mass objects would appear to require a relatively fast
migration of the type discussed here followed by a slowing down or
stopping near their current orbital locations.  The fate of a giant
protocore at the end of its runaway episode(s) if it has not been
brought close to the primary, would be a slow, type~II migration,
together with a possible mass growth, along the lines already studied
e.\ g. by \citet{n00} followed possibly by stopping inside a
magnetospheric cavity.

An alternative scenario could be envisaged to account for the
properties of EGPs.  In this, the disks in which they formed would
have been massive enough to sustain a succession of runaway episodes
typically up to the central tenth of an astronomical unit where hot
``Jupiters'' with $M\sin i\sim 0.3$~$M_J$, roughly corresponding to
the most favorable mass for runaway, are found to accumulate (or
alternatively protocores that massive could be constituted in situ by
the accumulation of lower mass bodies brought there by type I
migration.)  Some of these hot proto-Saturns could then be involved in
outward runaway.
At the same time, gas accretion onto these cores would eventually
endow them with a mass sufficient to prevent any further runaway
migration episodes. These planets would then correspond to the massive
($M_p > 1-2$~$M_J$) extrasolar planets, which are found further out
from their host stars than the hot Jupiters.

\begin{acknowledgements}
Computational resources were available at the CGCV Grenoble, and are
gratefully acknowledged.
\end{acknowledgements}

\appendix

\section{Corotation torque on a migrating planet for a given specific vorticity profile}
\label{apA}

\begin{figure}
\plotone{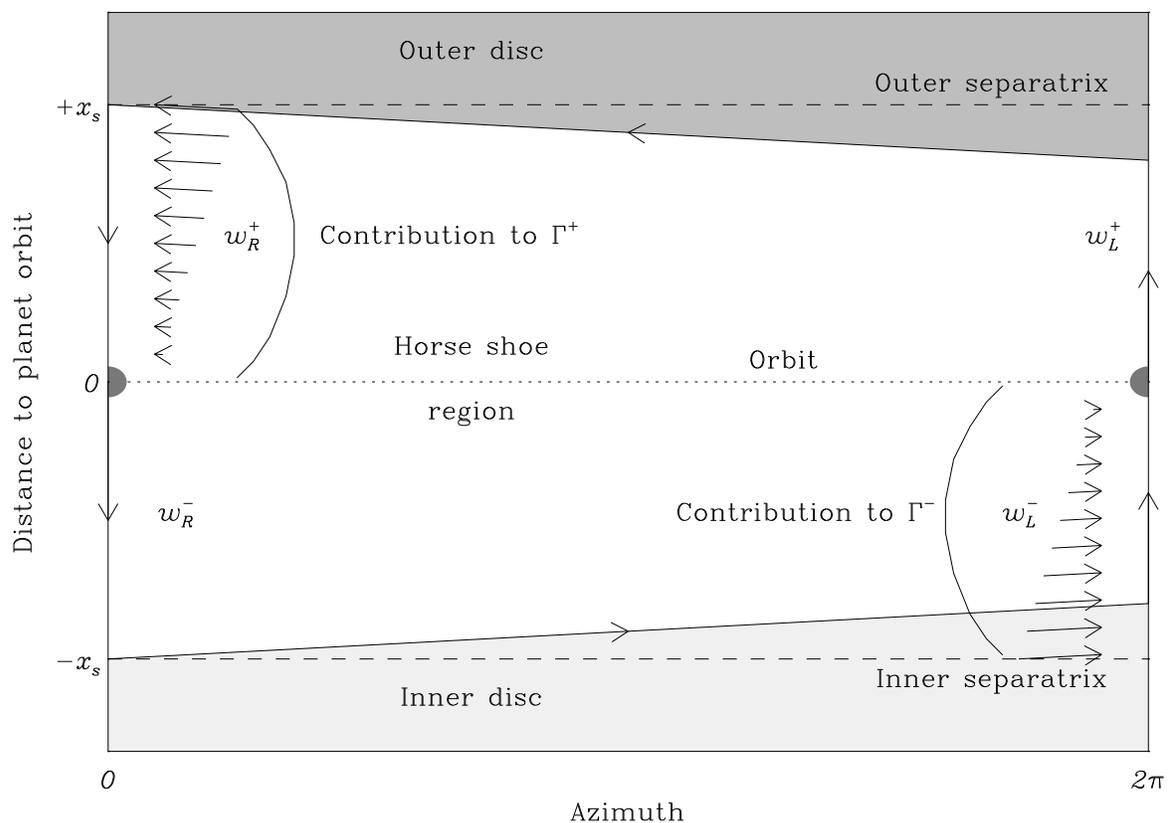}
\caption{Sketch of the flow for the case $\dot a<0$.
    We see that over the whole interval $[-x_s,+x_s]$, one has
    $w_R(x)=w_R(-x)$. On the other hand, one cannot write
    $w_L(-x)=w_L(x)$ for any $x$ in $[-x_s,+x_s]$, since fluid
    elements from the inner disk (light shaded region) contribute to
    the negative part of the corotation torque ($\Gamma^-$), and
    therefore bring into the coorbital region new material, with {\em
    a priori} an arbitrary specific vorticity.  The consequence of
    this distinction is that one has to evaluate the coorbital mass
    deficit from the upstream side of the coorbital depletion (i.\
    e. in this case from the mass flow across the inner separatrix.)}
\label{fig:appendix}
\end{figure}

\begin{figure}
\plotone{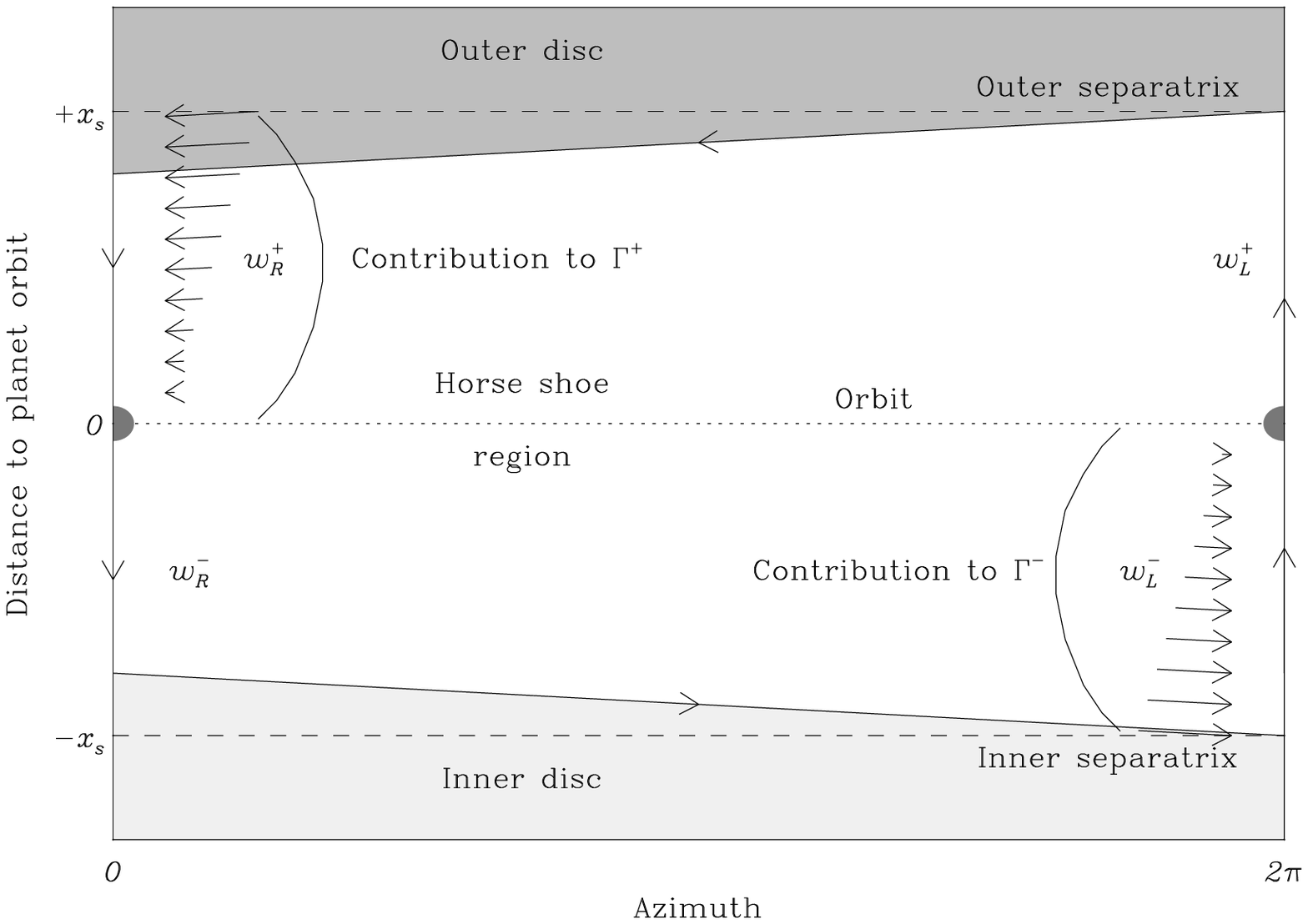}
\caption{Same as Fig.~\ref{fig:appendix} but with $\dot a>0$.
    We see that over the whole interval $[-x_s,+x_s]$, one has
    $w_L(x)=w_L(-x)$. On the other hand, one cannot write
    $w_R(x)=w_R(-x)$ for any $x$ in $[-x_s,+x_s]$, as new material
    from the outer disk is involved in the corotation torque (heavy
    shaded region), and has {\em a priori} an arbitrary specific
    vorticity. \label{fig:appendix2}}
\end{figure}

We transform Eq.~(\ref{eq:tqinter}) using the fact that a coorbital
fluid element does not undergo any radial drift (we assume a low
viscosity) between two successive close encounters and we write:
\begin{equation}
\label{eq:cons}
w_L^-(x,t)=w_R^-[x+a(t)-a(t-\tau),t-\tau],
\end{equation}
where we use the fact that the distance $r=x+a(t-\tau)$ of the fluid
element to the central object, just after the $R$-close encounter at
time~$t-\tau$, is the same as the distance $r=x+a(t)$, just before the
$L$-close encounter at time $t$.  We now assume that the profile
$w_R(x,t)$ is independent of $t$ for any $x$ in $[-x_s,+x_s]$, which
corresponds to assuming that either migration is steady state, or that
it is slow enough that the surface density profile responds much
faster than $|x_s/\dot a|$.  We can therefore write:
\begin{eqnarray}
w_L^-(x,t)&=&w_R^-(x,t)+[a(t)-a(t-\tau)]\frac{\partial w_R^-}{\partial x}\nonumber\\
&&+\frac 12[a(t)-a(t-\tau)]^2\frac{\partial^2 w_R^-}{\partial x^2}.
\end{eqnarray}
A Taylor expansion in~$\tau$ of this expression yields:
\begin{eqnarray}
w_L^-(x,t)&=&w_R^-(x,t)+\tau\dot a\left(\frac{\partial w_R^-}{\partial x}+
\frac 12\tau\dot a\frac{\partial^2 w_R^-}{\partial x^2}\right)\nonumber\\
&&-\frac 12\tau^2\ddot a\frac{\partial w_R^-}{\partial x}.\label{eq:inter2}
\end{eqnarray}
The second term of the bracket, an upper limit of which is
$\sim(1/2)\dot a\tau(\partial w_R^-/\partial x)/H$, is negligible
compared to the first one, as long as the migration remains slow.
Eqs.~(\ref{eq:tqinter}) and~(\ref{eq:inter2}) yield:
\begin{eqnarray}
\Gamma&=\displaystyle
16\pi B_p^2a^2&\left\{\dot a\left(x_sw_R^-(-x_s)-\int_{-x_s}^0
w_R^-(x)dx\right)\right.\nonumber\\
&&+\left.\frac{\pi a\ddot a}{2|A_p|}[w_R^-(0)-w_R^-(-x_s)]\right\}.\label{eq:tqf}
\end{eqnarray}
An order of magnitude estimate of the last term (in $\ddot a$) of
Eq.~(\ref{eq:tqf}) can be given if one assumes e.\ g. a quadratic
dependency of $w_R(x)$ on $x$: $w_R(x)\propto x^2$, in which case one
has: $w_R(-x_s)-w_R(0)=3\,\delta m\,/(16\pi ax_sB_p)$.
Eq.~(\ref{eq:tqf}) can then be rewritten as:
\begin{equation}
\Gamma=2B_pa\,\delta m\, \dot a
-\frac{3\pi a^2B_p}{2|A_p|x_s}\,\delta m\,\ddot a,
\end{equation}

which is similar to Eq.~(\ref{eq:tqexp}), except for a factor $3/2$ in
the $\ddot a$ term, which we had claimed to be given in order of
magnitude only.  

%
%

%
\end{document}